\UseRawInputEncoding
\documentclass[preprint,           
               showpacs,            
               nopreprintnumbers,     
               aps,                 
               prd,          	    
               letterpaper,             
               nofootinbib,         
               tightenlines,        
               floats,floatfix,      
               showkeys
               ]{revtex4-1}

\usepackage{graphicx}
\usepackage{dcolumn}
\usepackage{bm}
\usepackage{amsmath}
\usepackage{amsfonts,amssymb}
\usepackage{color}
\usepackage{enumerate}

\usepackage{subfig}
\usepackage{multirow,tabularx, booktabs}
\usepackage{placeins}

\usepackage{xcolor}
\usepackage{enumerate}

\def\arcsinh{\mathop{\rm arcsinh}\nolimits}
\definecolor{mygreen}{rgb}{0.0, 0.6, 0.0}

\begin{document}

\title{Enhancing Gravitational Lens Study with Deep Learning: A Study on Effects of Dropout Regularization}

\author{Juan J. Ancona-Flores$^{1}$}
\email{jancona16@alumnos.uaq.mx}

\author{A. Hern\'andez-Almada$^{1}$}
\email{ahalmada@uaq.mx}

\author{Ver\'onica Motta$^{2}$} 
\email{veronica.motta@uv.cl}

\affiliation{$^{1}$Facultad de Ingenier\'ia, Universidad Aut\'onoma de Quer\'etaro, Centro Universitario Cerro de las Campanas, 76010, Santiago de Quer\'etaro, M\'exico.}
\affiliation{$^2$Instituto de F\'isica y Astronom\'ia, Universidad de Valpara\'iso, Avda. Gran Breta\~na 1111, Valpara\'iso, Chile.}

\begin{abstract}
Strong gravitational lensing provides valuable insights into the mass distribution of galaxies and the nature of dark matter. However, its modeling is computationally demanding due to the large volume of strong lensing observations. In this work, we explore the application of Convolutional Neural Networks to infer physical parameters from simulated galaxy-galaxy lens systems, described by the Singular Isothermal Ellipsoid (SIE) profile for the galaxy lens. We construct a dataset of 76,396 synthetic lensing images derived from the China Space Station Telescope catalog and employ it to train a modified CNN model, based on AlexNet architecture, to predict four key SIE parameters, Einstein radius, axis ratio and ellipticity components. We analyze the network performance under three distinct dropout configurations to quantify their influence on generalization and parameter inference accuracy. The results indicate that the incorporation of dropout is critical for enhancing the precision and robustness of the estimated parameters, as demonstrated using a 4-fold cross-validation procedure. When dropout tools are included we obtain yields coefficients of determination up to $R^2 \sim 0.96$ for most SIE parameters and mean Peak Signal-to-Noise Ratios of up to $\sim 37$ dB. Relative to the configuration without dropout, the use of dropout reduces the relative errors in the inferred SIE parameters by approximately $60-76\%$, resulting in errors of at most $\sim 9\%$ at the 90\% confidence level for the majority of parameters. These findings highlight the potential of deep learning approaches to enable scalable, computationally efficient, and high-precision modeling of strong gravitational lensing systems.
\end{abstract}

\date{\today}
\maketitle

\section{Introduction}
A gravitational lens is any massive object, such as a galaxy, galaxy cluster, or quasar, whose gravitational influence results in the curvature of space-time, thereby altering the trajectory of light from more distant objects. When the projected mass within the Einstein radius becomes sufficient to produce multiple images, the observed phenomenon is known as a strong gravitational lens (SGL). When the alignment observer-lens-source is perfect, and the source is extended the resulting image is a ring known as an Einstein ring \cite{schneiderGL:2006, schneiderEAC:2006, hezavehDLSS:2016, meneghettiIGL:2021, shajibSLG:2024b}. Such phenomena impose significant constraints on the projected mass of the lensed object, aligning with the predictions of Einstein's theory of general relativity. Thus, strong lensing functions as a vital tool for probing dark matter within the central regions of halos or for determining the gradient of the inner mass density profile of galaxies \cite{TreuSLG:2010, shajibSLG:2024b}. Moreover, strong lenses facilitate the examination of sources with high redshift as a result of their ability to magnify up to 30 times the background source \cite{natarajanSLGC:2024a}. Grillo et al. \cite{grilloCPS:2008} proposed a methodology to estimate cosmological parameters using Strong Gravitational Systems (SGS). This method leveraged the correlation between the Einstein radius and the central stellar velocity dispersion, assuming an isothermal distribution of the total density for the lensed galaxy.
However, the problem with applying this method is the mass-sheet degeneracy and the deviations from isothermality and orbital isotropy that can cause the observed velocity dispersion to
differ by up to $20\%$ \cite{shajibSLG:2024b}. Solving this degeneracy requires spatially resolved kinematics, which might not be readily available for the thousands of lensed systems expected in future surveys. Thus, the method proposed by \cite{grilloCPS:2008} is a first approximation for such a large sample of systems. Another application is its use to estimate the Hubble constant $H_0$ by analyzing time delays between multiple quasar images of the source \cite{chaeCLASS:2003, birrerTDCM:2024, shajibSLG:2024b}. However, this method requires investing in months or years of photometric follow-up observations. 

Advanced and sophisticated telescopes, including Euclid \cite{Euclid:2025, laureijsEuclidDS:2011}, the Very Large Telescope \cite{enardVLT:1991}, the James Webb Space Telescope (JWST) \cite{mcelwainJWST:2023, gardnerJWT:2006} are already providing unprecedented data. Furthermore, next generations facilities such as the Simonyi telescope at the Vera C. Rubin Observatory \cite{biancoVCR:2021}, the Extremely Large Telescope (ELT) \cite{hookELT:2009, palleGES:2025} and the Chinese Survey Space Telescope (CSST) \cite{gongIntroductionChineseSpace:2026a} are expected to begin operations in the coming years. Together, these instruments will capture images and collect extensive datasets of approximately $\sim 10^5$ \cite{CollettGGSL:2015} to enhance our understanding of SGS. Some known large galaxy-galaxy lens surveys include the Cosmic Lens All-Sky Survey (CLASS) \cite{myersCLASS:2003, browneCLASS:2003a}, the COSMOS Survey \cite{scovilleCOSMOS:2007}, and the Sloan Lens ACS Survey (SLACS) \cite{boltonSLACS:2006, BoltonSLACSV:2008}. Such an extensive data collection, comprising more than 100,000 images, poses a challenge to traditional analytical methods, which may require substantial processing time. Conventional lens modeling techniques attempt to address the non-linear inverse problem of reconstructing the brightness of lensed sources while simultaneously modeling the gravitational potential of lenses.

Subsequent to the detection of lenses, it is imperative to develop a model that accurately depicts the total mass distribution, thereby assisting in the elimination of potential false-positive outcomes. Currently, Monte Carlo Markov Chain (MCMC) techniques have been deployed for system modeling; however, these methods are becoming inadequate given the increasing volume of data \cite{fowlieNSC:2020}. To mitigate this issue, artificial intelligence tools such as Convolutional Neural Networks (CNNs) are highly valued due to their proven success in stellar image classification, thereby becoming the burgeoning technique for lens search and identification. CNNs were first introduced in 1988 when Zhang \cite{zhangSNN:1992} proposed the first two-dimensional convolutional neural network, known as the shift-invariant artificial neural network. The inaugural application of CNNs in astronomical classification involved spectrum classification in the SDSS, as documented by H\'ala's \cite{HalaSCNN:2014}. Petrillo et al. \cite{petrilloFSG:2017} pioneered the use of a CNN-based morphological classification method to identify strong gravitational lenses in the 225 Kilo Degree Survey. Hezaveh showcased the proficiency of CNNs in predicting parameter values of mass models using preprocessed high-resolution lens images \cite{hezavehFAA:2017}. This team further expanded their research by presenting a CNN specifically designed for modeling high-resolution lens images. The Highly Optimized Lensing Investigations of Supernovae, Microlensing Objects, and Kinematics of Ellipticals and Spirals (HOLISMOKES) team developed a CNN aimed at modeling strongly lensed galaxy images \cite{schuldtHOLIVE:2021}, crafted to predict the five parameters of the SIE mass model. Warren R. Morningstar et al. \cite{morningstarDRG:2019} introduced a machine learning approach to reconstructing undistorted images of background sources in strongly lensed systems. Additionally, Bayesian Neural Networks (Bayesian NN), a probabilistic variant of deep neural networks, have exhibited great success in the extraction of highly abstract information from complex image data. This capability is exemplified in the work of Ji Won Park et al. \cite{parkLGL:2021}, who used simulated lens time delays to determine the Hubble constant, and Pearson et al. \cite{pearsonSLM:2021}, who trained an approximate Bayesian CNN to predict the parameters of the mass profile. More recently, the Euclid Collaboration \cite{euclidQDR:2025} designed a Bayesian NN to model gravitational lenses automatically and efficiently. With LEMON, key parameters of the lens mass profile, such as the Einstein radius, are estimated in addition to the parameters of the light distribution of the lens galaxy, along with its uncertainties. LEMON was applied to simulated lenses (with characteristics similar to those captured by the Euclid telescope) and also to real lenses. Further works such as those by R. Parlange et al. \cite{TomasVGraViT:2025} dives into transformer models to detect strong gravitational lenses from sets from HOLISMOKES VI and SuGOHI X. 

Due to the efficacy of CNNs in analyzing SGS, this study provides a comprehensive examination of CNN performance in predicting the physical parameters associated with strong lensing. Specifically, our research emphasizes the influence of CNN architecture and regularization strategies, with particular attention to the incorporation of dropout layers on the accuracy and robustness of parameter estimation.  Dropout, a widely used technique introduced by N. Srivastava et al.\cite{srivastavaDSW:2014}, plays a crucial role in mitigating overfitting in deep learning models, thereby enhancing their generalization capabilities. Its functionality consists on randomly turning off a set of neurons during each training iteration, preventing the units from depending on each other and enabling the network to learn more complex structures. Dropout technique is more detailed in Section \ref{section:cnn}. In contexts such as gravitational lensing, where the recovery of model parameters such as ellipticity and external shear is acutely sensitive to minute variations in image features, the optimization of dropout placement and rate is crucial in enhancing prediction quality. Consequently, our objective is to determine the optimal configuration and integration of dropout layers within a CNN model to enable precise and reliable estimation of the parameters of the single isothermal ellipsoid (SIE) model. Our detailed analysis involves comparing the complete image set against the discrepancies observed between simulated images and those modeled based on CNN-predicted values. The synthetic dataset comprises 76,396 images of simulated galaxy-galaxy lens systems derived from the China Space Station Telescope (CSST) catalog, generated in a noise-free environment. In our analysis, the dataset was partitioned into 70,000 images for the k-fold cross-validation process, in which each fold utilized 75\% of the images for training and 25\% for validation. Additionally, the 6,393 remaining images were used in testing purposes. A model based on the AlexNet architecture \cite{AlexNet:2012} was selected because actual deep architectures are excessive and computationally expensive to process the millions of images that the great telescopes are going to get. The modifications made to the network enable a smaller architecture compared to current ones, facilitating rapid predictions. The addition of a convolutional layer to the central block enhances data extraction depth and mitigates linearity. The model predicts four parameters from the SIE lens model: complex ellipticity components $\epsilon_x$ and $\epsilon_y$, axial ratio $f$ and Einstein radius $\theta_E$ (see Section \ref{section:sgl} for their definitions). The predictive accuracy of the model is assessed by measuring the relative error between the actual and predicted values. Through this investigation, our aim is to advance the interpretability and effectiveness of deep learning methodologies in modeling gravitational lenses through the validation of the dropout as a critical tool for the reliability of predictions and the use of a lighter architecture designed for the massive data.

The manuscript is organized as follows. Sections \ref{section:sgl} and \ref{section:cnn} introduce the subject matter and articulate the principal aim of the paper. These sections provide a theoretical foundation concerning gravitational lensing and succinctly expound on CNNs. Section \ref{section:method} explores the synthetic data used in this investigation, elaborating on the factors considered in the methodological approach, and offers a comprehensive overview of the evaluation metrics used to assess the performance of the model and analyze the relative errors. Section \ref{section:results} provides a summary of the performance of the model, elucidates the relative errors associated with the predictions, and discusses these findings. Finally, Section \ref{section:conclusions} presents the conclusions derived from the study and outlines recommendations for future research.

\section{Gravitational lensing background}
\label{section:sgl}
\subsection{Strong lensing theory}
\begin{figure*}
\centering
\includegraphics[width = 0.5\linewidth]{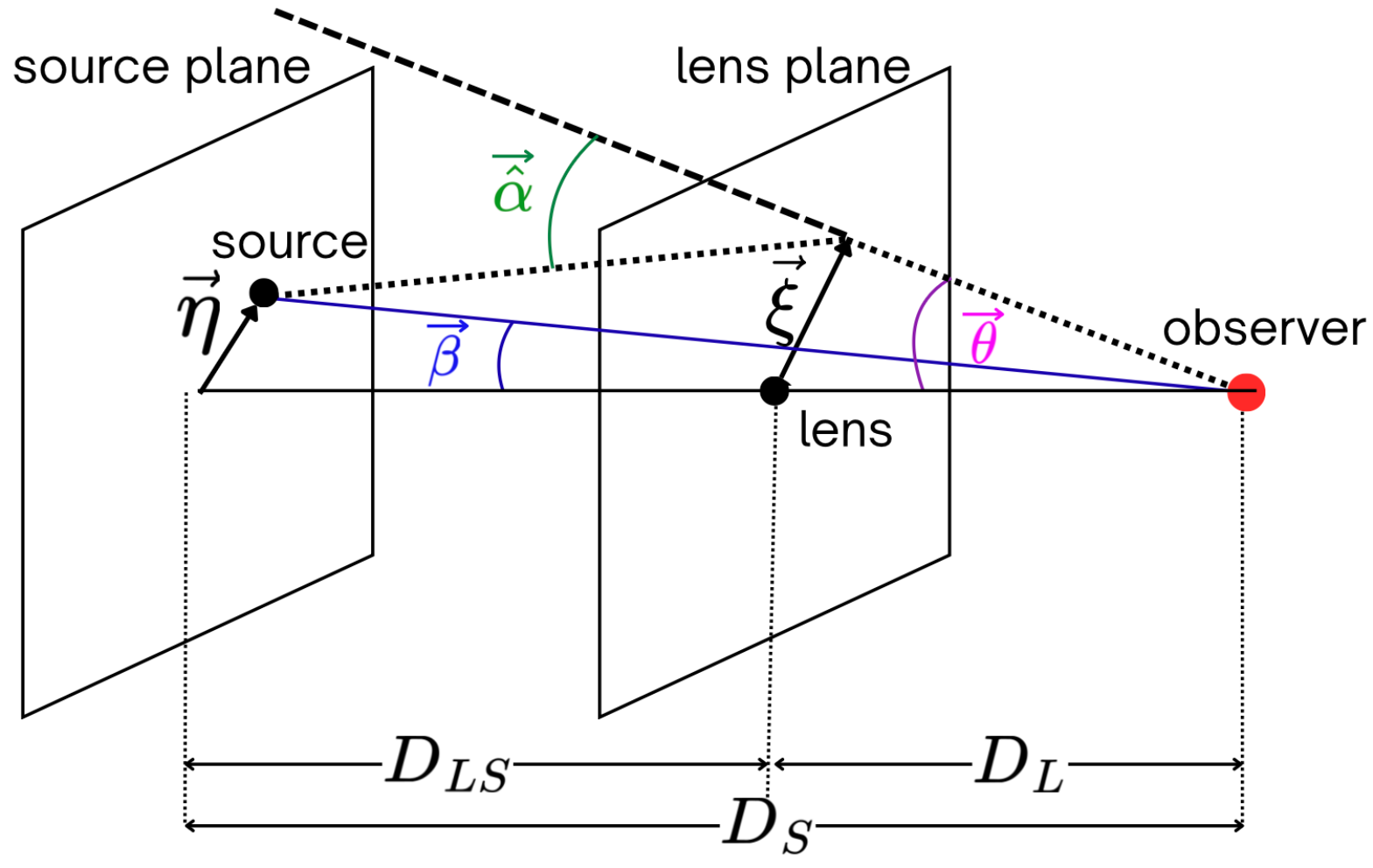}
\caption{Sketch not drawn to scale, illustrates the parameters involved in a typical gravitational lensing system. $\vec{\beta}$ represents the real position of the source, $\vec{\theta}$ represents the apparent position of the source and $\vec{\hat{\alpha}}$ represents the deflection angle. }
\label{fig:lensscheme}
\end{figure*}

A representative scenario in gravitational lensing is depicted in Figure \ref{fig:lensscheme}, where a mass concentration is observed to deflect light rays emanating from a source located at a distance $D_{S}$. In the absence of additional deflectors near the line of sight, and assuming that the extent of the deflecting mass along the line-of-sight is considerably smaller than both $D_L$ and the distance $D_{LS}$ from the deflector to the source, light rays are considered to be deflected in a plane (thin lens hypothesis). The magnitude and orientation of this curvature are characterized by the deflection angle $\vec{\hat{\alpha}}$, which depends on the mass distribution of the deflector and the impact vector of the light ray \cite{schneiderGL:2006}. The real position of the source relative to the observed position of the images in the sky is therefore delineated by the {\it lens equation}, which outlines a transformation from the lens plane to the source plane for any mass distribution.

For a source positioned in the source plane $\vec{\eta} = \vec{\beta}D_S$ with a real angular position $\vec{\beta}$, the light rays that arrive at the lens plane at a distance  $\vec{\xi} = \vec{\theta}D_L$ from the optical axis (defined by the lens-observer line) are deflected by the angle $\vec{\hat{\alpha}}$ before reaching an observer. As a result of the deflection, the observer perceives images of the source at the apparent angular positions $\vec{\theta _i}$, with $i$ representing each image. If $\vec{\theta}$, $\vec{\beta}$, and $\vec{\hat{\alpha}}$ are small angles (of the order of arcseconds), considering the reduced deflection angle by

\begin{equation}
\vec{\alpha}(\vec{\theta}) \equiv \dfrac{D_{LS}}{D_S}\vec{\hat{\alpha}},
\label{eq:deflectonangle}
\end{equation}
where $D_{LS}$ is the angular diameter distance between the lens and the source, the lens equation can be written as:
\begin{equation}
\vec{\beta} = \vec{\theta} - \vec{\alpha}(\vec{\theta}).
\label{eq:lensequation2}
\end{equation}
One important consequence of gravitational lensing is image distortion, which can be described by the Jacobian matrix as
\begin{equation}
A \equiv \dfrac{\partial \vec{\beta}}{\partial \vec{\theta}} = \left(\delta_{ij} - \dfrac{\partial^2 \hat{\Psi}(\vec{\theta})}{\partial \theta_i\partial \theta_j}\right),
\label{eq:distortion}
\end{equation}
where $\hat{\Psi}$ represents the effective lensing potential. By segregating the isotropic component (associated to the mass of the main deflector) from the Jacobian and extracting its trace-less segment, we can derive the {\it shear tensor}, which encapsulates the external perturbation, thereby characterizing the distortions of background sources. The expression for the shear tensor is
\begin{equation}
\Gamma = \left(\dfrac{1}{2}\text{tr}A\cdot I - A\right)_{ij} =
\begin{pmatrix}
\gamma_1 & \gamma_2\\
\gamma_2 & -\gamma_1
\end{pmatrix},
\label{eq:shear}
\end{equation}
where $\gamma_1 = \gamma\cos 2\phi$ and $\gamma_2 = \gamma\sin 2\phi$. 

\subsection{SIE model parametrization}
In order to construct realistic models of gravitational lens systems, we must employ mass distributions defined by elliptical isodensity contours. In this research, we focus on the surface density of the singular isothermal Ellipsoid (SIE) model \cite{kormannIEG:1994, tessoreEPL:2015, asadaIIE:2003, meneghettiIGL:2021}. The SIE model can be understood as a natural generalization of the singular isothermal sphere (SIS), which assumes spherical symmetry and an isothermal velocity distribution. In the SIS model, three-dimensional mass density ($\rho(\vec{r})$) follows:
\begin{equation}
    \rho(\vec{r}) = \dfrac{\sigma_v^2}{2\pi Gr^2},
    \label{eq:densityprofile}
\end{equation}
where $\sigma_v^2$ represents the line-of-sight velocity dispersion of the lens galaxy and $r$ is given by $r = \sqrt{\xi^2 + z^2}$, which represents the spherical radial coordinate. To account for deviations from spherical symmetry, the SIE model introduces ellipticity in the projected mass distribution by replacing the circular radial coordinate in the lens plane with an elliptical one. Following \cite{kormannIEG:1994}, this is achieved through the transformation $\xi \rightarrow \sqrt{\xi_1^2 + f^2\xi_2^2}$. Under this prescription, the projected density profile  and the corresponding surface mass density ($\Sigma(\vec{\xi})$) become
\begin{align}
    \rho(\vec{r}) = \dfrac{\sigma_v^2}{2\pi Gr^2}, \label{eq:siedensityprofile}\\
     \Sigma(\vec{\xi}) = \dfrac{\sigma_v^2}{2G}\dfrac{\sqrt{f}}{\sqrt{\xi_1^2 + f^2\xi_2^2}}.
    \label{eq:surfacedensity}
\end{align}
In this context, $f$ refers to the axis ratio of the ellipses, specified within the interval $0 < f \leq 1$. The SIE model profile is fundamentally defined by the line-sight velocity dispersion of the lens galaxy, $\sigma_v$. This motivation is key, as $\sigma_v$ serves as an observational input to alternatively estimate the galaxy mass. Consequently, the dimensionless surface mass density or convergence $\kappa(\vec{\theta})$ is directly related to this dispersion. Given the angular coordinates $\vec{\theta} = (\theta_1, \theta_2)$, the convergence is given by:
\begin{equation}
\kappa(\theta_1, \theta_2) = \dfrac{\theta_E\sqrt{f}}{2\sqrt{\theta_1^2+f^2\theta_2^2}}\,.
\label{eq:surfdensprof}
\end{equation}
The expression delineates the density of the projected surface mass within the lens. Within this formulation, given a perfect alignment source-lens-observer, $\theta_E$ denotes the Einstein radius, which is defined by
\begin{equation}
\theta_E = 4\pi\dfrac{\sigma_v^2}{c^2}\dfrac{D_{\text{LS}}}{D_\text{S}},
\label{eq:einsteinradii}
\end{equation}
where $c$ denotes the speed of light. Using polar coordinates $(r, \phi)$, where $\phi$ is the polar angle measured from the semi-major axis ($\hat{e}_1$), and taking into account the relationship delineated in $\Delta \Psi = 2\kappa$, where $\Psi$ represents the lensing potential for the SIE model, which can be expressed as
\begin{equation}
\Psi(x, \phi) = x\dfrac{\sqrt{f}}{f'}\left[\sin\phi\arcsin(f'\sin\phi) + \cos\phi\arcsinh \left( \dfrac{f'}{f}\cos\phi \right)\right],
\end{equation}
where $f' = \sqrt{1 - f^2}$ and characterizes the shape of equipotential surfaces and tends to unit at the circular boundary \cite{kormannIEG:1994, schneiderGL:2006, meneghettiIGL:2021}. The deflection angle relevant to SIE model can be expressed as follows
\begin{equation}
\vec{\alpha}(\vec{x}) = \dfrac{\sqrt{f}}{f'}\left[ \arcsinh\left(\dfrac{f'}{f}\cos\phi \right)\hat{e}_1+ \arcsin(f'\sin\phi)\hat{e}_2 \right]\,,
\end{equation}
 where $\hat{e}_1$ and $\hat{e}_2$ denote the unit vectors along the principle axes and the deflection angle is related to the potential by$\vec{\nabla}\Psi = \vec{\alpha}$.
In addition to the free parameters of the model, which include the central coordinates and the position angle $\varphi$ between the semi-major axis of the mass ellipse of the lens and the axis $x$ of the lens plane, the ellipticity is typically parameterized by the eccentricity moduli
\begin{align}
\epsilon_{l,x} = \dfrac{1-f}{1+f}\cos2\varphi, \label{eq:ellipticity1} \\
\epsilon_{l,y} = \dfrac{1-f}{1+f}\sin2\varphi, \label{eq:ellipticity2}
\end{align}
which are continuously defined according to $-1 < \epsilon_i < 1$, thus removing the periodic boundaries associated with $\phi$ and addressing the discontinuity in $f=0$ as noted in \cite{etheringtonAGG:2022}.

\subsection{Light model parametrization}

We assume that the lens and source light follow the elliptical S\'ersic model distribution, allowing a flexible but physically motivated description of galaxy morphology while keeping the number of free parameters manageable for lens modeling. The S\'ersic profile for the surface brightness is defined by the following expression.
\begin{equation}
I(x,y) = I_e\exp\left[-b_{n_s}\left\{\left(\dfrac{\sqrt{x^2 + y^2/f^2}}{R_e}\right)^{1/n}-1\right\}\right]\,, \label{eq:sersiclens}
\end{equation}
where $I_e$ denotes the surface brightness at the half-light radius $R_e$, and $b_{n_s}$ represents a constant related to the Sérsic index $n_s$ by $\Gamma(2n_s) = 2\gamma(2n_s, b_{n_s})$ \cite{sersicIAI:1963, sersicAGA:1968, R1mlawCiotti:1999, cardoneLPS:2004}. 

\section{Convolutional Neural Networks theory}
\label{section:cnn}

Convolutional neural networks, although analogous to traditional feedforward neural networks, are characterized by their ability to extract hierarchical features through convolutional processes, thus reducing computational expenses \cite{liSCN:2022, alzubaidiRDL:2021}. CNNs have several advantages over conventional neural networks. Firstly, they reduce the number of parameters and enhance convergence by establishing neuron connections with smaller subsets of neurons from the previous layer. Secondly, the method of weight-sharing decreases parameters by distributing weights across various connections. Lastly, CNNs incorporate downsampling, which reduces the input dimensionality, making them suitable for smaller input sizes. 

\begin{figure*}
\centering
\includegraphics[width = 0.9\textwidth]{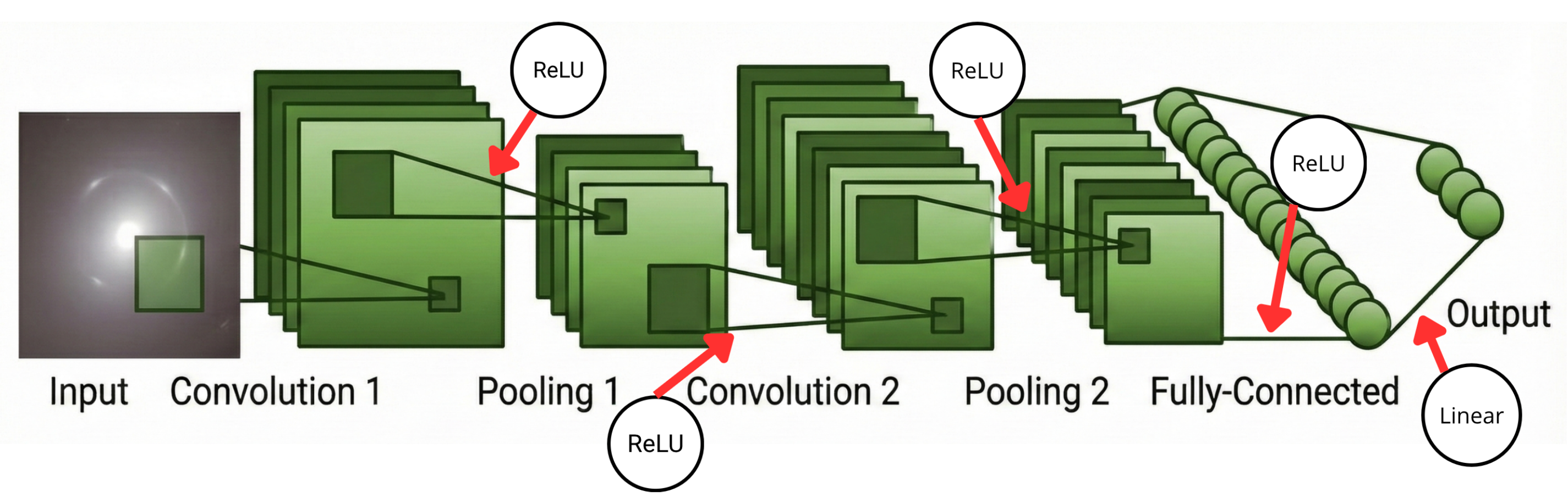}
\caption{Illustrative architecture of a CNN model. The design incorporates convolution layers for the extraction of local features, which are subsequently reduced in spatial dimensions through pooling layers. The processed features are subsequently managed by fully connected layers to yield the final output, corresponding either to the prediction of physical parameters or to classification tasks. An important component of NNs are the activation function, which manage the activation of the neurons in each layer.}
\label{fig:cnnschema}
\end{figure*}

The CNN architecture comprises four main components: convolutional layers, pooling layers, activation functions, and fully connected layers, as depicted in Figure \ref{fig:cnnschema}. Convolutional layers utilize filters on the input data to identify patterns. The pooling layers perform simplification by downsampling along spatial dimensions, thus reducing parameter loads in the activation process. Their core function is to reduce the spatial dimensions of the activation maps while preserving critical features, which helps to reduce computational demands, control overfitting, and achieve invariant representations of small translations. Activation functions are integral to the expression of complex features, acting similarly to the neuronal model of the human brain by determining which information propagates to the succeeding neuron. The Rectified Linear Unit (ReLU) is prevalently used among activation functions \cite{nairReLU:2010, glorotDSRNNRELU:2011, AlexNet:2012}; it assists the model in discerning non-linear image patterns. If the input $x$ is negative, its function value is 0, while if $x$ is non-negative, its function value equals $x$ itself. The ReLU function is written as
\begin{equation}
    ReLU(x) = 
    \left\{ 
    \begin{matrix}
    0 & \text{if} \ x \le 0 \\
    x & \text{if} \ x > 0
    \end{matrix}
    \right. = \max(0,x)
\end{equation}
Although the term CNN was assigned in the 1980s, it was Yann LeCun \cite{lecunBP:1989} who pioneered the effective application of the backpropagation method for CNNs aimed at handwritten zip code recognition, coining the term {\it convolution} in the process. Dropout is a significant machine learning algorithm introduced in the conference on Neural Information Processing Systems (NIPS) 2012 \cite{baldiUD:2013} to train neural networks, which works by randomly {\it dropping out} neurons during training to prevent feature detector co-adaptation \footnote{Co-adaptation occurs when several characteristic detectors "agree" and work well only if they are all together. Every neuron stops learning useful signs by themselves and become a "parasite" of the rest. Dropout breaks that group behavior: turn off random neurons during training to force them to learn robust and independent things}. Dropout facilitates the training of numerous network configurations within reasonable timescales \cite{hintonINN:2012}, alleviates overfitting, and offers an effective approximation to combine exponentially many unique NN structures \cite{srivastavaDSW:2014, salehinDropout:2023}. In considering a NN with hidden layers $L$, where $\vec{z}^l$ is the input vector for each layer, $\vec{y}^l$ represents the output vector from layer $l$, and $\vec{w}^l$ signifies the weights in layer $l$, the dropout feedforward algorithm undergoes redefinition,
\begin{align}
r_j^l &\sim \text{Bernoulli}(p), \\
\bar{\vec{y}}^l &= \vec{r}^l\odot\vec{y}^l, \\
z_i^{(l+1)} &= \vec{w}_i^{(l+1)}\cdot \bar{\vec{y}}_i^l + b_i^{(l+1)},\\
y_i^{(l+1)} &= f(z_i^{(l+1)}),
\end{align}
where $\vec{r}^l$ represents a vector of independent Bernoulli random variables, each associated with probability $p$. The output of that layer is multiplied in elements by $\vec{r}^l$ to produce thinned outputs $\bar{\vec{y}}^l$ \cite{srivastavaDSW:2014}. The probability of dropping out depends on the architecture and size of the dataset, and its study is crucial to obtain promising results and reduce overfitting. Normally, the rate at which the dropout is selected is between 20\% and 50\%. In the subsequent section, we examine three scenarios concerning the influence of dropout on CNN outputs.

\section{Methodology}
\label{section:method}

\subsection{Simulated sample from China Space Station Telescope (CSST)}
\begin{figure*}
    \centering
    \includegraphics[width=14.1cm]{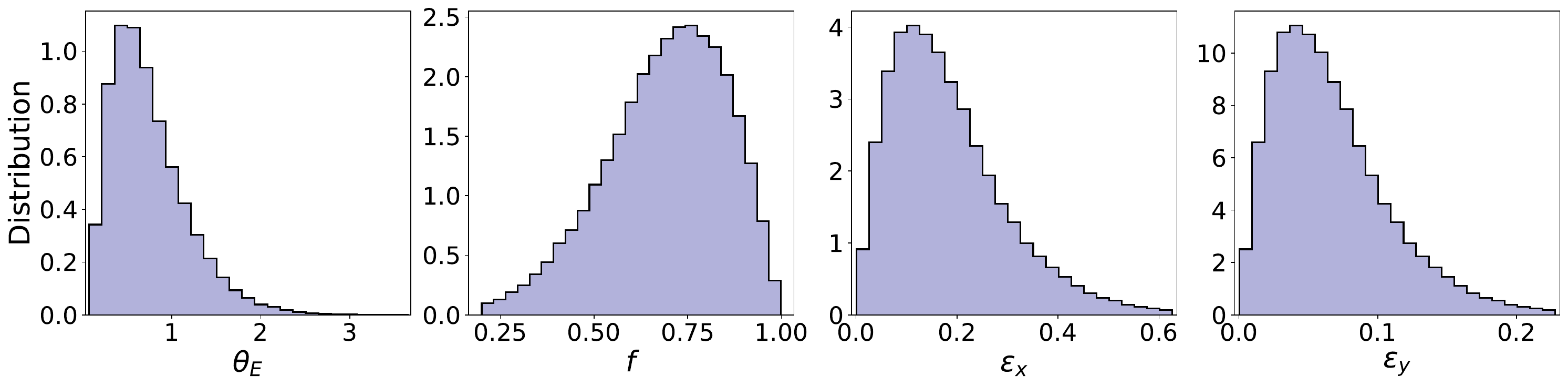}
    \caption{Marginal distributions of the training parameters of the SIE lens model.}
    \label{fig:paramdist}
\end{figure*}

\begin{figure*}
\centering
\includegraphics[width = \textwidth]{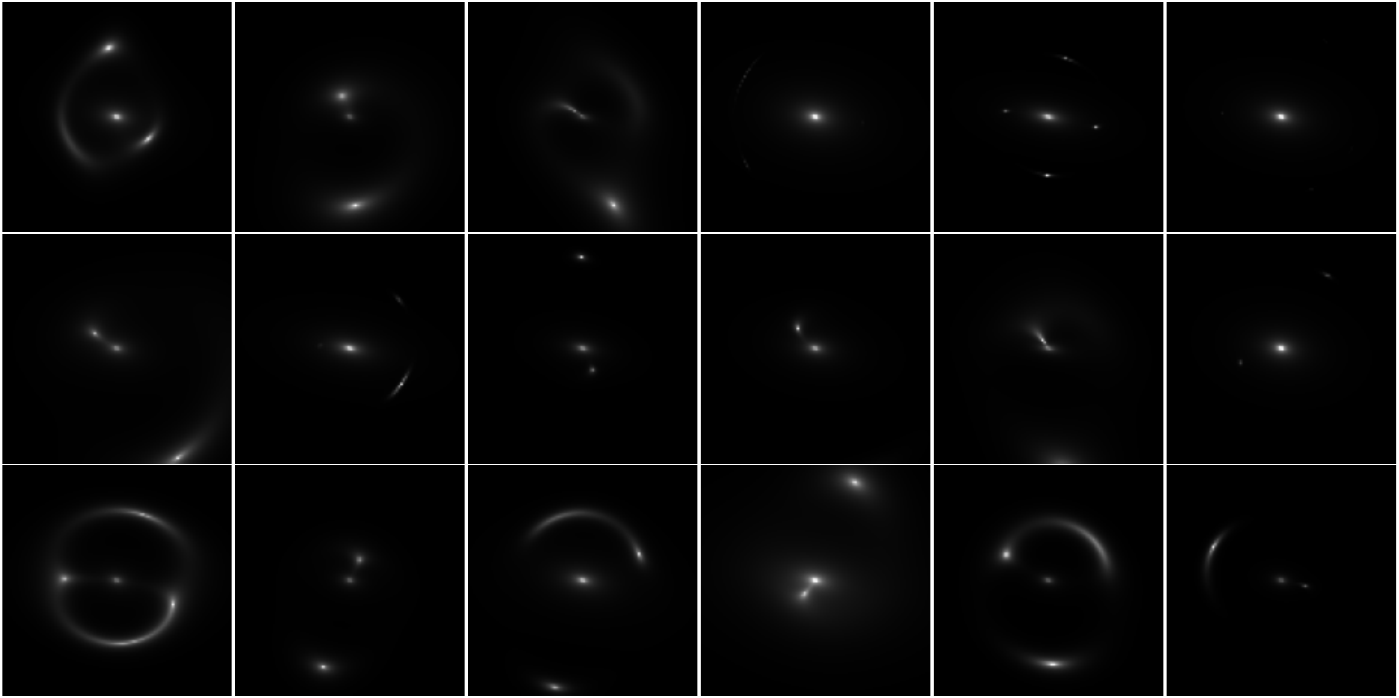}
\caption{Random examples of synthetic images of galaxy-galaxy lens systems used in training with a resolution of 0.06 arcsec/pixel. The $100\times100$ pixel$^2$ grid ensures a field of view that captures the position of the lensed images. The lens systems were simulated considering the SIE lens model and using the CSST catalogue \cite{CaoCSST:2024}.}
\label{fig:imagesexample}
\end{figure*}

The data set used to train, validate and test the CNN model consists of 76,396 images, each with dimensions of $100 \times 100$ pixels with 1 channel and a resolution of $0.06$ arcseconds/pixel, based on the configuration of strong lens systems presented by CSST \cite{CaoCSST:2024}, corresponding to the single band Wide-Field sample. The images are generated using the Lenstronomy package, in the Python environment \cite{birrerLMG:2018}, in which the source is modeled by the S\'ersic model characterized by the space parameter $\Theta_s=(R_e, \epsilon_{s,x}, \epsilon_{s,y}, x, y)$, where $x$ and $y$ represent the position of the source with respect to the lens plane and setting the S\'ersic index $n_s = 2.0$. This value of the S\'ersic index was chosen because it represents an intermediate value between distinct galaxy populations ($n_s = 1.0$ for disk galaxies and $n_s = 4.0$ for elliptical galaxies). Furthermore, S. Mukherjee et al. \cite{mukherjeeSEAGE:2018} found that the precise choice of source parameters for lens modelling is of secondary importance and does not significantly bias lens modeling. As shown in equations \ref{eq:ellipticity1} and \ref{eq:ellipticity2}, the components of the ellipticities were computed knowing the axis ratio from configurations of the CSST sample \cite{CaoCSST:2024}. For this work, the position angle (p.a.) used to compute the components of the ellipticities for the lens ($\epsilon_{l,x}, \epsilon_{l,y}$) was arbitrary set at $\text{p.a.} = 10$~deg from North, while the p.a. used to calculate the ellipticities components for the source where extracted from the CSST sample. The lens is characterized by the SIE model for the mass distribution with parameters $\Theta_l=(\theta_E, \epsilon_{l,x}, \epsilon_{l,y})$ and the S\'ersic model for the brightness distributions used for the source and lens. For the following Sections, the notation ($\epsilon_x$,$\epsilon_y$) will be adopted to refer to the ellipticities of the lens. Figure \ref{fig:paramdist} shows the distribution for each parameter used in training issues. The distributions for $\theta_E$ and ellipticity components ($\epsilon_x$, $\epsilon_y$) exhibit a clear positive skew, while axis ratio ($f$) shows a concentration toward higher values. For $\theta_E$, the values are concentrated between [0.07, 3.65]. For $f$, the values can be found between [0.20, 1.00]. The ellipticities components ($\epsilon_x$, $\epsilon_y$) can be found between [0.00, 0.63] and [0.00, 0.23], respectively. Figure \ref{fig:imagesexample} shows several random examples of synthetic images used in the analysis. To train the model, the dataset was partitioned into 70,000 images for the k-fold cross-validation process\footnote{k-fold cross-validation is a statistical technique used to evaluate the generalization capacity of a model and prevent overfitting. It consists of dividing the data set into $k$ folds; the model is trained using $k-1$ of those groups and is validated with the remaining group that was left out \cite{StoneMCrossval:2018}.} and 6,396 images for independent testing. Withing the 4-fold cross-validation scheme, each fold utilized 75\% of the images (52,500) for training and 25\% (17,500) for validation.

\subsection{CCN}

The primary objective is to train an AlexNet-modified CNN (described later), to predict the four characteristic parameters of the SIE model ($\theta_E$, $f$, $\epsilon_x$, $\epsilon_y$) for lens modeling, assessing the performance of the CNN model under three conditions in which the dropout layers were altered. As mentioned previously, The dataset was partitioned into 70,000 images for 4-fold cross-validation for training and validation and 6,396 images for independent testing. In the 4-fold cross-validation scheme, each fold used 75\% of the images for training and 25\% for validation. A crucial element of the data set is the binary format used to store images, namely TFRecords, which efficiently stores data sequences, optimized for use with TensorFlow \cite{tfrecord:2024}. This format is ideal for large data sets, as it allows for more compact storage and faster data reading during model training, even if the data do not fit completely in memory. This format facilitates the systematic labeling and organization of images, allowing a single file to be partitioned into training, validation, and test samples. Furthermore, the format's support for grayscale images streamlines preprocessing procedures.

The CNN model used in this research is based on the AlexNet architecture, which incorporates modifications in both the number of layers and neurons. This architecture, introduced by Alex Krizhevskii et al. in 2012 \cite{AlexNet:2012, BallesterPGA:2016, zhangANLNVGG:2021}, achieved significant success in the ImageNet Large Scale Visual Recognition Challenge of 2012, an effort that evaluates algorithmic performance in object detection and image classification. AlexNet achieved an error rate of 15.3\% in the classification of large-scale images. AlexNet expands the foundational concepts of LeNet and applies the core principles of CNNs to a more extensive and deep network. Furthermore, AlexNet integrates the ReLU activation function, dropout, and local response normalization (LRN) for the first time within a CNN. For the purposes of this preliminary study, this model was selected among various available models due to its notable performance, as evidenced by works such as those of Perreault et al. \cite{perreaultlUPE:2017}, which demonstrate its ability to accurately estimate lens parameters.

\begin{figure*}
\centering
\includegraphics[width = 0.55\textwidth]{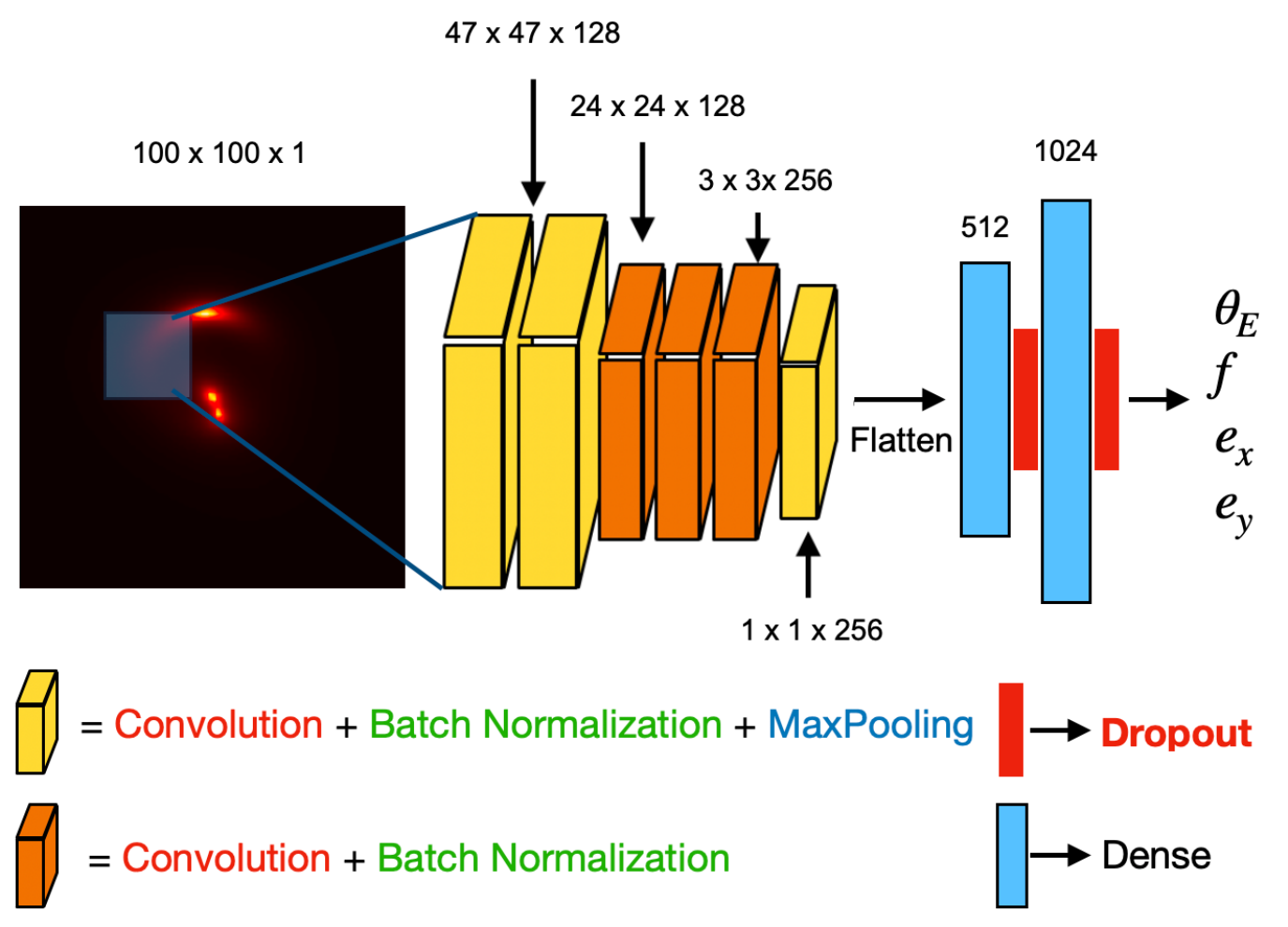}
\caption{The CNN that we use consists of a series of convolution groups interspersed with max pooling layers and batch normalization terminating in two dense layers with two layers of dropout followed by an output layer as shown in the figure. }
\label{fig:alexnetmodified}
\end{figure*}

The architecture of the AlexNet-based CNN model is shown in Figure \ref{fig:alexnetmodified}. The model comprises six blocks preceding the fully connected layers. The initial blocks consist of a convolutional layer, followed by batch normalization and maximum-pooling layers. The first convolutional layer applies 128 filters of size $7\times 7$ with a stride of 2, thereby reducing the spatial resolution. The second convolutional layer utilizes 256 filters of size $5\times 5$ with a stride of 3, facilitating the extraction of mid-scale features. The deep convolutional block, consisting of four convolutional layers with 256 filters of size $1\times1$, captures hierarchical representations and enhances the depth of the network without considerably increasing the number of parameters. This block concludes with a Max-Pooling layer to further reduce spatial dimensionality. The flattened feature maps are processed by two dense layers consisting of 512 and 1024 neurons, respectively. The investigation places significant emphasis on dropout layers, examining changes in dropout rates across three scenarios. Ultimately, the model concludes with a dense layer possessed of linear activation and yielding four outputs corresponding to the parameters of interest. The primordial improvement that presents the model is the BatchNormalization layers, which stabilize training, allow higher learning rates, and reduce sensitivity to initialization. Another advantage of the architecture used in this work is the inclusion of an additional convolutional layer. This layer enhances the model’s ability to extract more details from the images. Additionally, the architecture undergoes a significant change with the implementation of filter factorization within the $1\times1$ convolutional layer block. This provides a better representation of the characteristics while keeping the computational cost. Because the ellipticity components ($\epsilon_x$, $\epsilon_y$) produce weaker variations in the images than other parameters of the lens model, we observed that the network tended to underadjust them. To mitigate this effect, we use a weighting scheme in the loss function defined by equation \ref{eq:weightmse}, assigning higher weights to the terms associated with $\epsilon_x$ and $\epsilon_y$. The weights for the Einstein radius and the axis ratio were 1.0 and for the ellipticity components 3.0, respectively. This strategy increases the contribution of these parameters to the gradient during training, stimulating the model to learn more sensitive representations of its morphological signatures.

The NAdam optimizer, an adaptation of the Adam optimizer that incorporates Nesterov momentum \cite{dozatINM:2016}, was used for weight refinement. Nesterov's Accelerated Gradient (NAG) is a first-order optimization technique that ensures a higher convergence rate in certain contexts compared to the traditional gradient descent method. This momentum distinguishes itself from the Stochastic Gradient Descent (SGD) optimizer by first taking a substantial step in the direction of the accumulated gradient, followed by the evaluation of the gradient's terminal position to perform necessary adjustments. NAG achieves a global convergence rate of $\mathcal{O}(1/T^2)$, marking an advance over the convergence rate of gradient descent $\mathcal{O}(1/T)$ \cite{sutskeverIIM:2013}. This refinement not only accelerates the convergence process but also enhances precision in minimizing the loss function. To mitigate overfitting, a ReduceLROnPlateau callback was integrated into the model, systematically reducing the learning rate and thus improving training metrics. This technique allows a dynamic adaptation of the learning rate: if the loss of validation does not show a significant improvement after a threshold of $n$ times (patience), the learning rate is reduced by a factor $\lambda$ \cite{cholletKeras:2015, GoodfellowDL:2016}. In this work, the factor for ReduceLROnPlateau was fixed in $\lambda = 0.5$. The chosen minimum learning rate was $\epsilon = 10^{-8}$. A significant experimental analysis involved modifying the dropout rates of two dropout layers, which were systematically tested under three conditions: initially using dropout rates of 20\% and 30\%, then uniformly applying a dropout rate of 20\% to both layers, and finally, disabling the dropout layers entirely. For further evaluation of the model, a test sample consisting of 6,396 images was selected, with the relative errors related to each prediction visualized.

\subsection{Evaluation metrics}
To assess and compare the performance of the model, the weighted Mean Squared Error (MSE) was selected as the loss function, setting the weights for the Einstein radius and axial ratio in 1.0 and the weights for the ellipticity components of the lens in 3.0, following the Multi-Task Learning approach proposed by Caruana (1997) \cite{caruanaMultitaskLearning:1997}, while the Mean Absolute Error (MAE) served as the evaluation metric \cite{ruppertElements:2004, BotchkarevMetrics:2019}. Furthermore, the coefficient of determination $R^2$ was calculated, providing an indication of the extent to which the variance in the observed values is explicable by the predictions of the model, expecting a value near 1 \cite{ChiccoR2:2021}. These metrics are articulated in the following expressions,
\begin{align}
{\rm MSE} =& \dfrac{1}{n}\sum_{i=1}^{n}\sum_{j=1}^{k}w_j(y_{i,j} - \bar{y}_{i,j})^2, \label{eq:weightmse}\\
{\rm MAE} =& \dfrac{1}{n}\sum_{i=1}^{n}|y_i - \bar{y}_i|, \\
R^2 =& 1 - \dfrac{\sum_{i=1}^{n}(y_i - \bar{y}_i)^2}{\sum_{i=1}^{n}(y_i - \hat{y})^2},
\end{align}
in which $y_i$ denotes the actual value, $\bar{y}_i$ represents the predicted value, and $\hat{y}$ signifies the mean of the actual values. Additionally, we assess the relative error inherent in the predictions by determining the residuals between the forecasted and actual values and subsequently plotting the actual values against the forecasted values \cite{HyndmanACCURACY:2006}. The relative error was evaluated employing the elementary definition of relative error
\begin{equation}
\epsilon_r = \left|\dfrac{\bar{y} - y}{\bar{y}}\right|.
\end{equation}
In order to evaluate the accuracy of the predictions the models on the test images, we computed the bias defined as:
\begin{equation}
    \mu = \text{median}(\vec{\bar{y}} - \vec{y}),
    \label{eq:bias}
\end{equation}
where $\vec{\bar{y}}$ corresponds to the predicted values array and $\vec{y}$ to the true values array. The scatter of the distribution can be obtained by computing using the 16th and the 84th percentiles of the distribution by the next definitions:
\begin{align}
    \sigma^{-} &= \text{median}(\vec{\bar{y}} - \vec{y}) - P_{16}(\vec{\bar{y}} - \vec{y}),\label{eq:loewrstd} \\ 
    \sigma^{+} &= P_{84}(\vec{\bar{y}} - \vec{y}) - \text{median}(\vec{\bar{y}} - \vec{y}). \label{eq:upperstd}
\end{align}
Furthermore, we determined the Normalized Median Absolute Deviation (NMAD), which is a robust statistic used to measure data dispersion, defined as the Median Absolute Deviation (MAD) scaled by a factor of approximately $k=1.48$ \cite{HuberRS:2011}. NMAD is defined as
\begin{equation}
    \text{NMAD} = 1.48 \ \text{median}[|(\vec{\bar{y}} - \vec{y}) - \mu|].
    \label{eq:NMAD}
\end{equation}
A critical examination to reinforce the findings involves simulating images with the predicted outputs of the CNN model in order to facilitate a comparative analysis using the PSNR and MSE between the images, allowing an objective assessment of the quality of the results. Taking into account a reference image $X$ and a test image $Y$ of size $M\times N$, the PSNR, measured in dB, is determined by $\text{MSE}_{\text{img}}$ between them using the following expression
\begin{equation}
    \text{PSNR}(X, Y) = 10\log_{10}(\text{MAX}_{\text{img}}^2/\text{MSE}_{\text{img}}(X, Y)),
    \label{eq:psnr}
\end{equation}
where 
\begin{equation}
    \text{MSE}_{\text{img}} = \dfrac{1}{MN}\sum_{i=1}^M\sum_{j=1}^N(X_{ij} - Y_{ij})^2,
    \label{eq:mseimages}
\end{equation}
and $\text{MAX}_{\text{img}}$ is the maximum value that a pixel can have in the image. A good value of PSNR is in a range between 30-40 dB. A PSNR value above 40 dB means a high-quality reconstruction of the original image \cite{sadykovaQAPSNR:2017, petrovicPSNR:2016}.

\section{Results}
\label{section:results}

\begin{figure*}
\centering
\includegraphics[width = 0.9\textwidth]{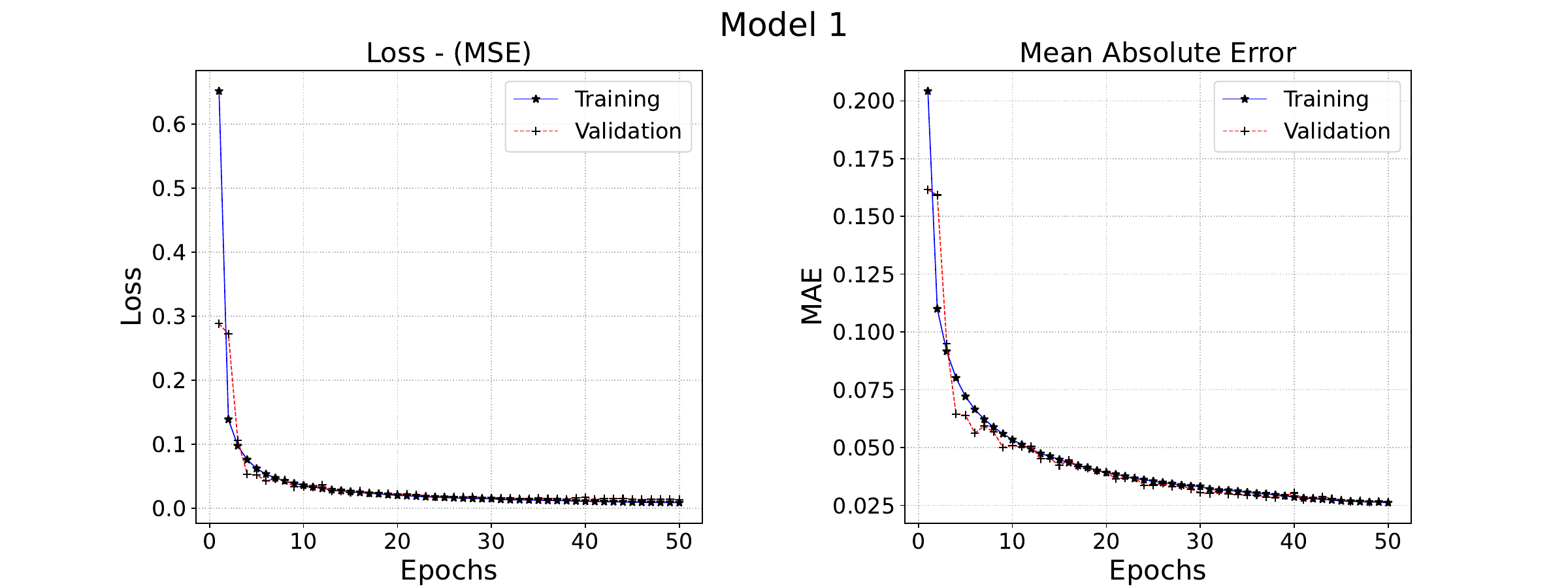}
\includegraphics[width = 0.9\textwidth]{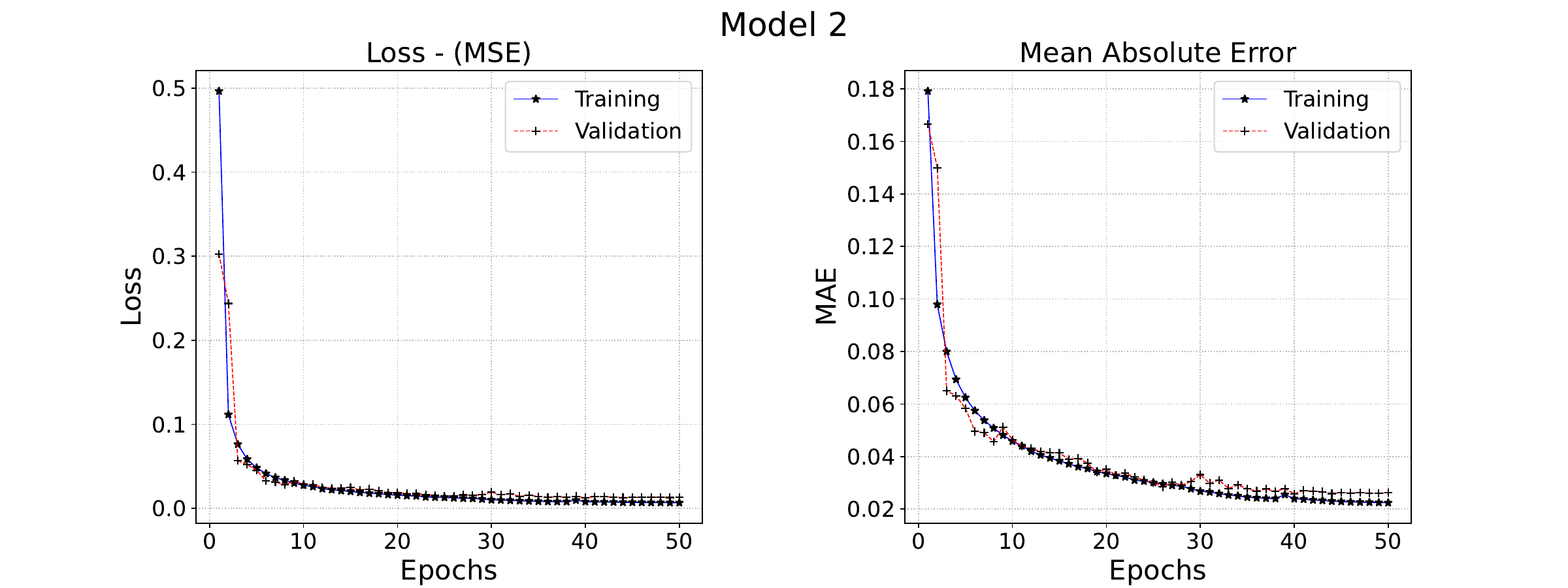}
\includegraphics[width = 0.9\textwidth]{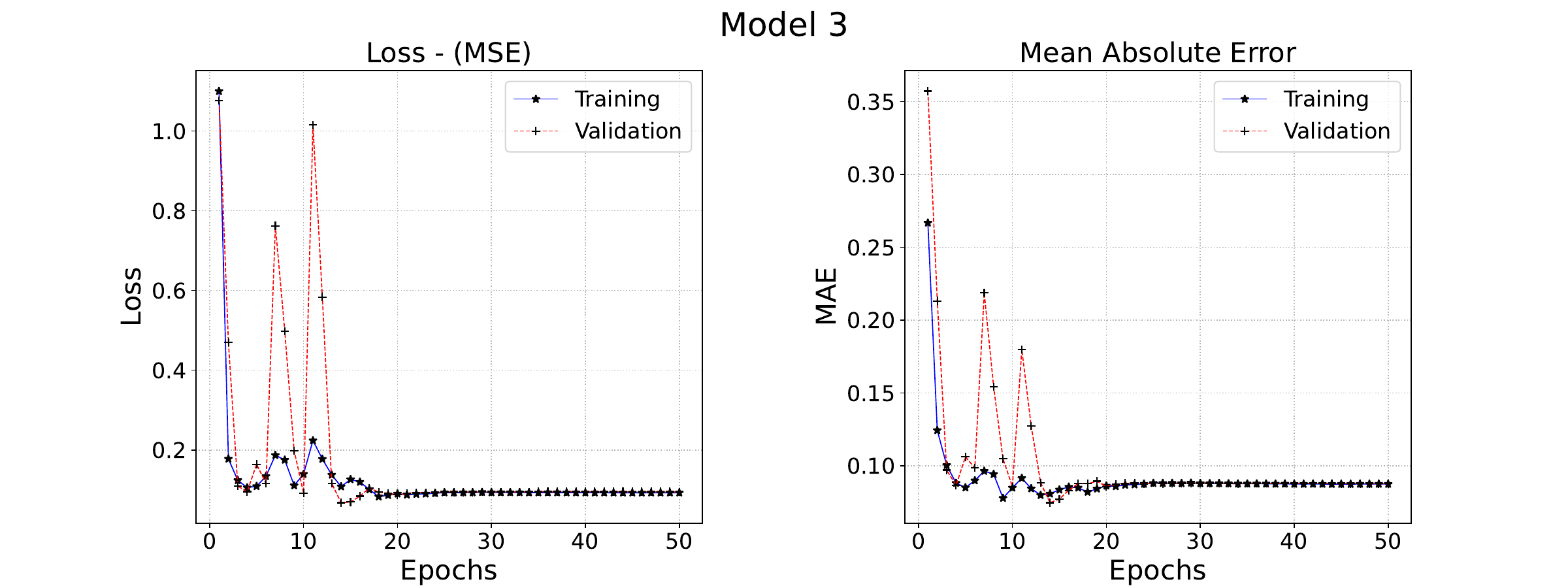}
\caption{The performance metrics of the three models that were analyzed during the training phase. 
The MSE average values and their standard deviation at the end of the training for Model 1, Model 2, and Model 3 were $0.008\pm0.001$, $0.006\pm0.001$, and $0.092\pm0.016$, respectively, while the corresponding MAE average values and their standard deviation were $0.026\pm0.001$, $0.022\pm0.001$, and $0.085\pm0.005$, respectively.}
\label{fig:metrics_cnn}
\end{figure*}

This section is devoted to analyzing the performance of the CNN model in predicting four parameters of the SIE model (lens galaxy). We note that the reported results correspond to the average values and their standard deviation obtained from a cross-validation procedure employing a 4-fold partitioning scheme. Specifically, the dataset was divided into four independent subsets. Figure \ref{fig:metrics_cnn} illustrates the evolution of loss and MAE functions with respect to the epochs, for which models 1 and 2 exhibit consistent behavior. Although both functions of model 3 display greater instability during the initial training phase, they eventually converge. Nevertheless, based on the minimum values attained by these metrics, as shown in Table \ref{table:coefr2}, model 3 achieves the highest (i.e., worst) loss and MAE values among the three CNN models.

\begin{figure*}
    \includegraphics[width=0.9\textwidth]{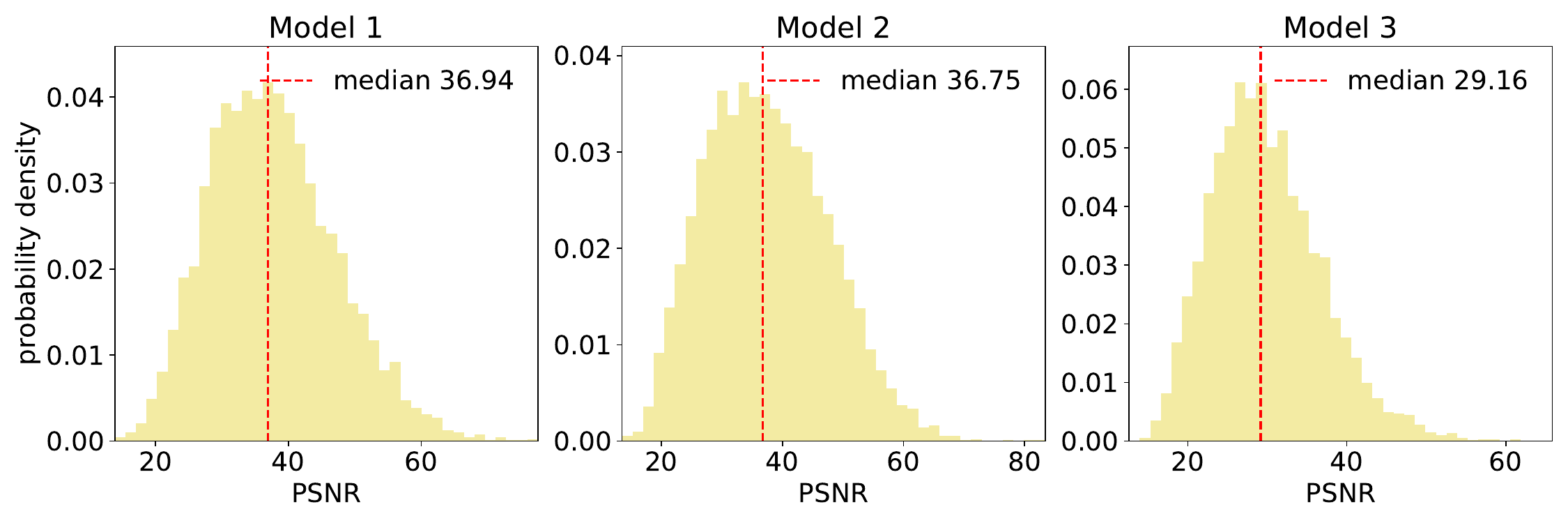}
    \caption{PSNR values for all the images used in test where the red dotted line indicates the median value. It can be seen that the predictions of models 1 and 2 generate images with high-quality compared to the third model.}
    \label{fig:psnrmeans}
\end{figure*}

\begin{figure*}
\includegraphics[width=0.9\textwidth]{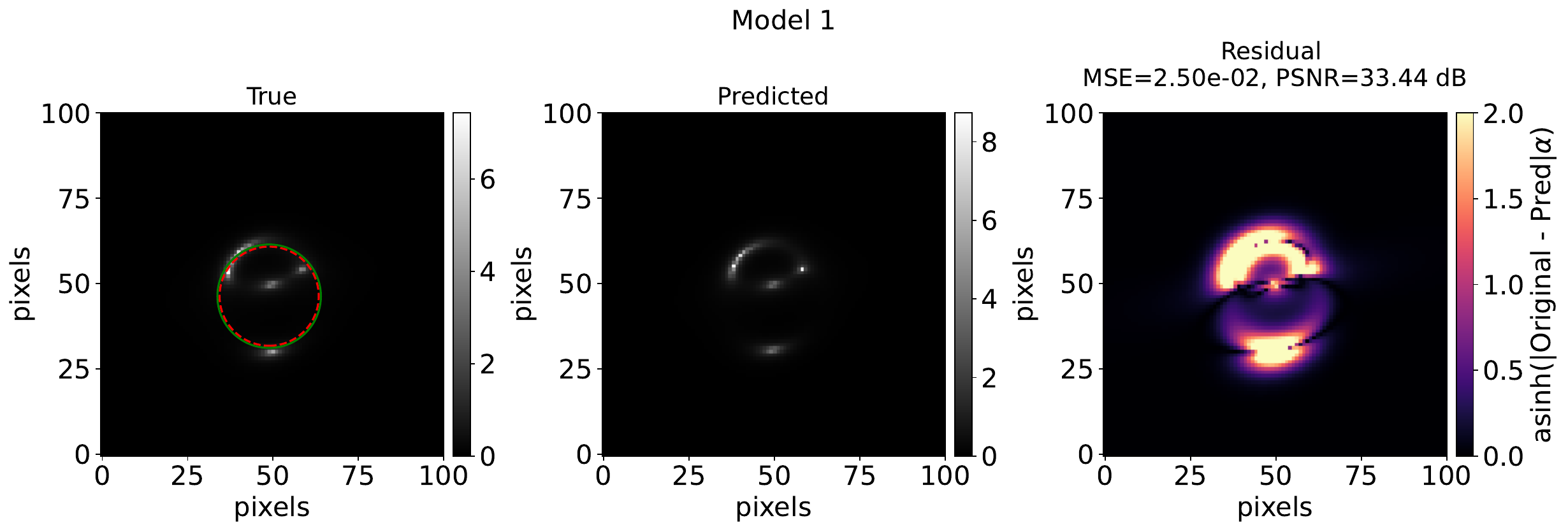}
\includegraphics[width=0.9\textwidth]{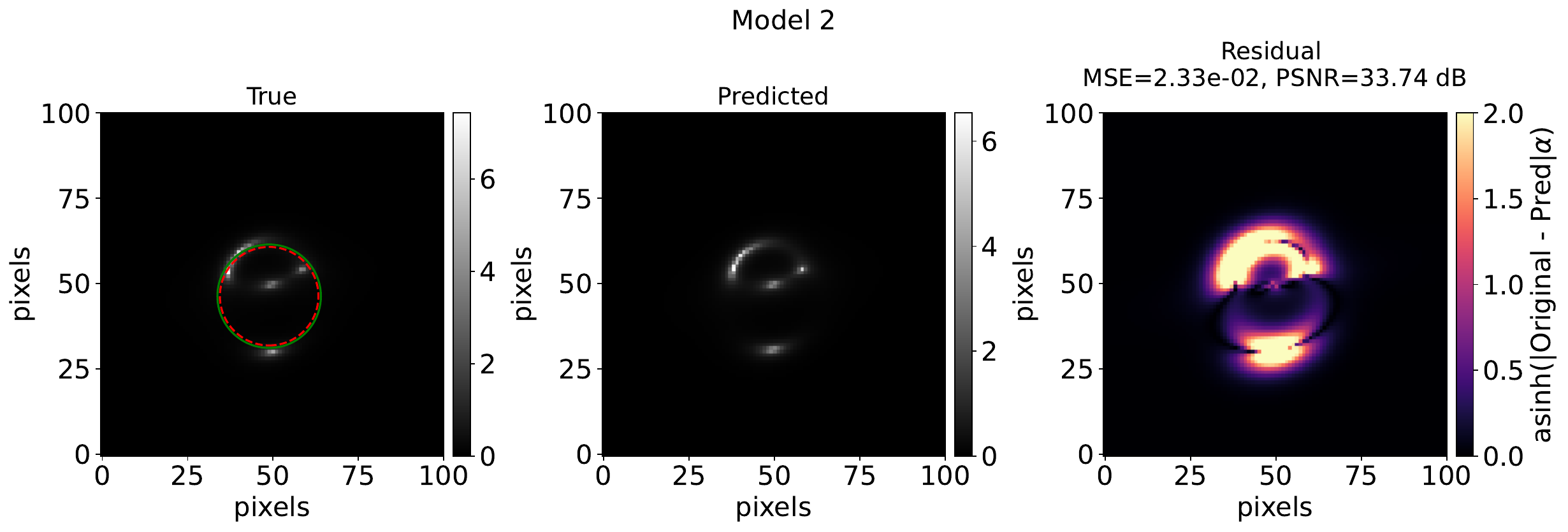}
\includegraphics[width=0.9\textwidth]{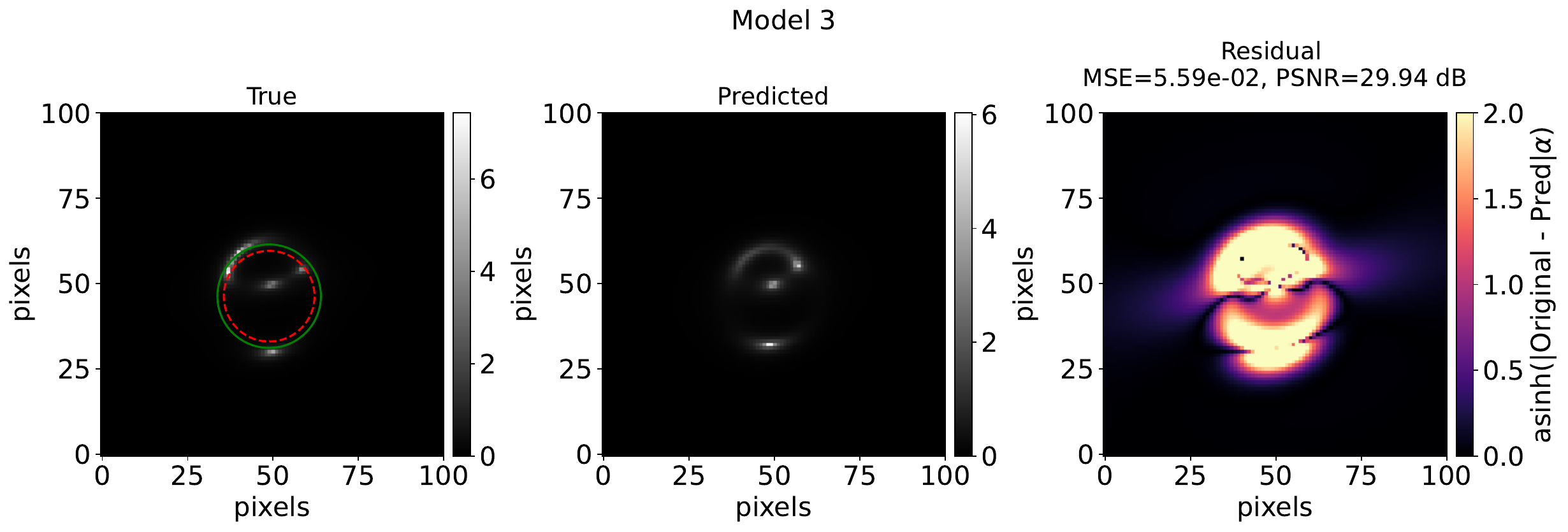}
\caption{Random example of a galaxy-galaxy lens systems. Left column shows the original synthetic image and the predicted synthetic image is shown on the middle column for the three CNN models. Right column is the distribution of the residual, difference between the original image and the predicted one. Furthermore, the Einstein ring $\theta_E$ is drawn using the predicted values (green circles) and the true values (red circles) of the lens system. The intensity on the residuals distribution is scaled using the $\arcsinh$ Lupton Transformation to show details of bright and weak areas. This transformation behaves linearly for weak signals and logarithmically for strong signals by setting $\alpha$; which in this case is given a value of $\alpha = 1/P_{95} =40$, related with the percentile 95 from the means of the residuals to avoid extreme atypical values in the display scale.} 
\label{fig:imgcomparative}
\end{figure*}

In the following, the results presented are obtained using the testing sample corresponding to 6396 images. Additionally, $R^2$ indicates that models that incorporate dropout exhibit superior predictive performance for SIE parameters (see Table \ref{table:coefr2}), achieving values in the range of 0.95 to 0.97, while model 3 achieves values comparatively lower (>0.56). Furthermore, Figure \ref{fig:psnrmeans} shows the distribution of the PSNR for the three models, in which the median value is denoted by a dotted red line. Based on this quantity, we find a good reconstruction of the images of the lensed source for model 1 and model 2 compared to model 3, obtaining similar results (median values of $\approx 37$) with dropout being considered. Figure \ref{fig:imgcomparative} displays a random example of a galaxy-galaxy lens system reconstructed using the true values (left panel) and the predicted values (middle panel). The right panel shows the residual distribution obtained as the difference between the true image and the predicted. As the intensity of the residual images is very small, we use a Lupton-type asinh transformation \cite{LuptonCCD:2004} that highlight the details of bright and weak areas by transforming weak signals linearly and strong signal logarithmically. We observe that the worst predicted reconstruction occurs when model 3 is used.

\begin{figure*}
\includegraphics[width = \textwidth]{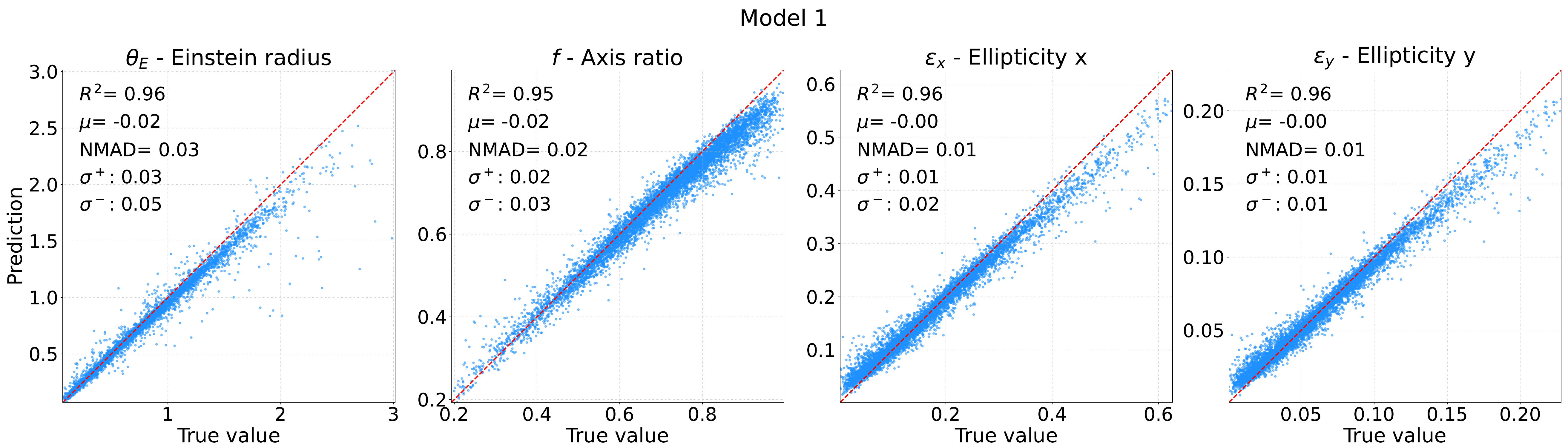}
\includegraphics[width = \textwidth]{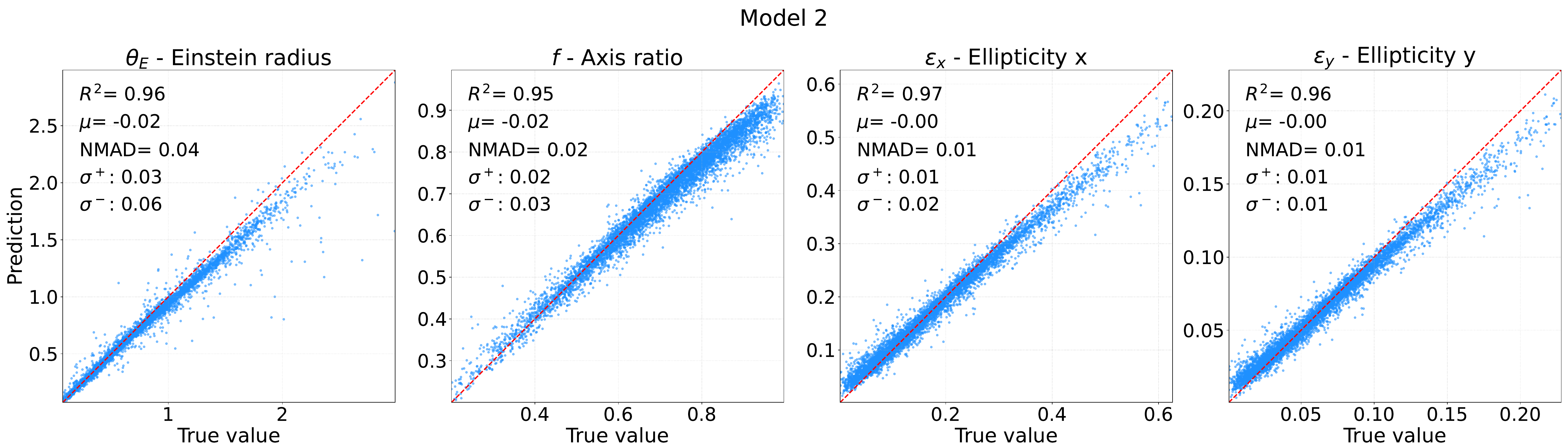}
\includegraphics[width = \textwidth]{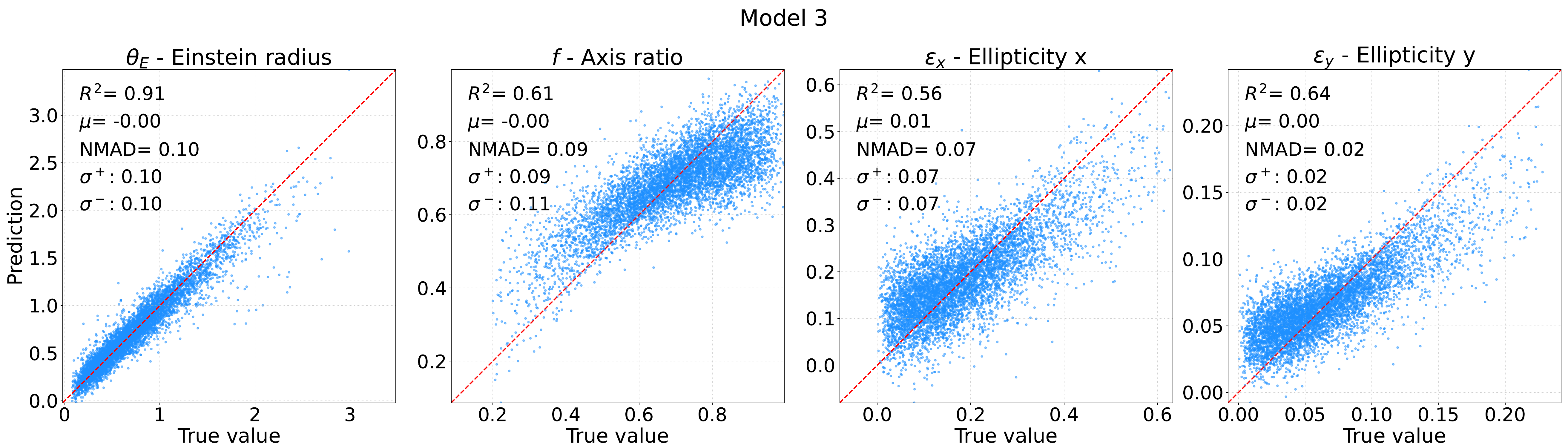}
\caption{Predicted vs. true parameters for the lens SIE model (Einstein radius, axis ratio, ellipticity components) for the three models. Models 1 and 2 show consistent high accuracy, while Model 3 has significant dispersion, especially in ellipticity and axis ratio predictions. The bias ($\mu$) is shown in the plots, along with the NMAD and dispersion limits. Predicted values are averages from the 4-fold cross-validation.}
\label{fig:scatter}
\end{figure*}

Figure \ref{fig:scatter} presents the scatter distribution between the true and predicted values of the SIE parameters. Exact predictions lie on the 45-degree diagonal (solid red line), and ideal predictions correspond to a distribution tightly aligned with this line. As anticipated, models 1 and 2 show broadly consistent results: they exhibit a higher predictive accuracy for low values of $\theta_E$, while a systematic bias becomes apparent for $\theta_E \gtrsim 1.2$. This can be associated with the values of $\theta_E$ utilized for training, as depicted in Figure \ref{fig:paramdist}. This figure illustrates a significant concentration of values within the range of $(0, 1.2)$. Similarly, the parameter $f$ displays a noticeable bias for $f \gtrsim 0.8$, where the models systematically underestimate the true values. For components $\varepsilon_x$ and $\varepsilon_y$, the bias is even more pronounced in the regimes $\varepsilon_x \gtrsim 0.4$ and $\varepsilon_y \gtrsim 0.12$, respectively. In contrast, model 3 shows the largest dispersion and bias among the three, clearly revealing the significant impact of dropout on the quality of the predictions. To assess the uncertainties of the models, we also computed the bias, the standard deviation and the NMAD, which serve as indicators of the accuracy of the predictions. The results of these metrics reveal that Models 1 and 2 exhibit a negligible systematic bias ($\mu \approx -0.02$) but extremely low NMAD values between 0.01 and 0.04, confirming that predictions are highly certain and present minimal dispersion with respect to the line of truth. In contrast, while Model 3 maintains a low average bias ($\mu \approx 0.00$), its high residual dispersion (NMAD $\approx 0.07-0.10$) indicates a regime of high variance that limits its reliability for individual parameter inference.

\begin{figure*}
\centering
\includegraphics[width=7cm]{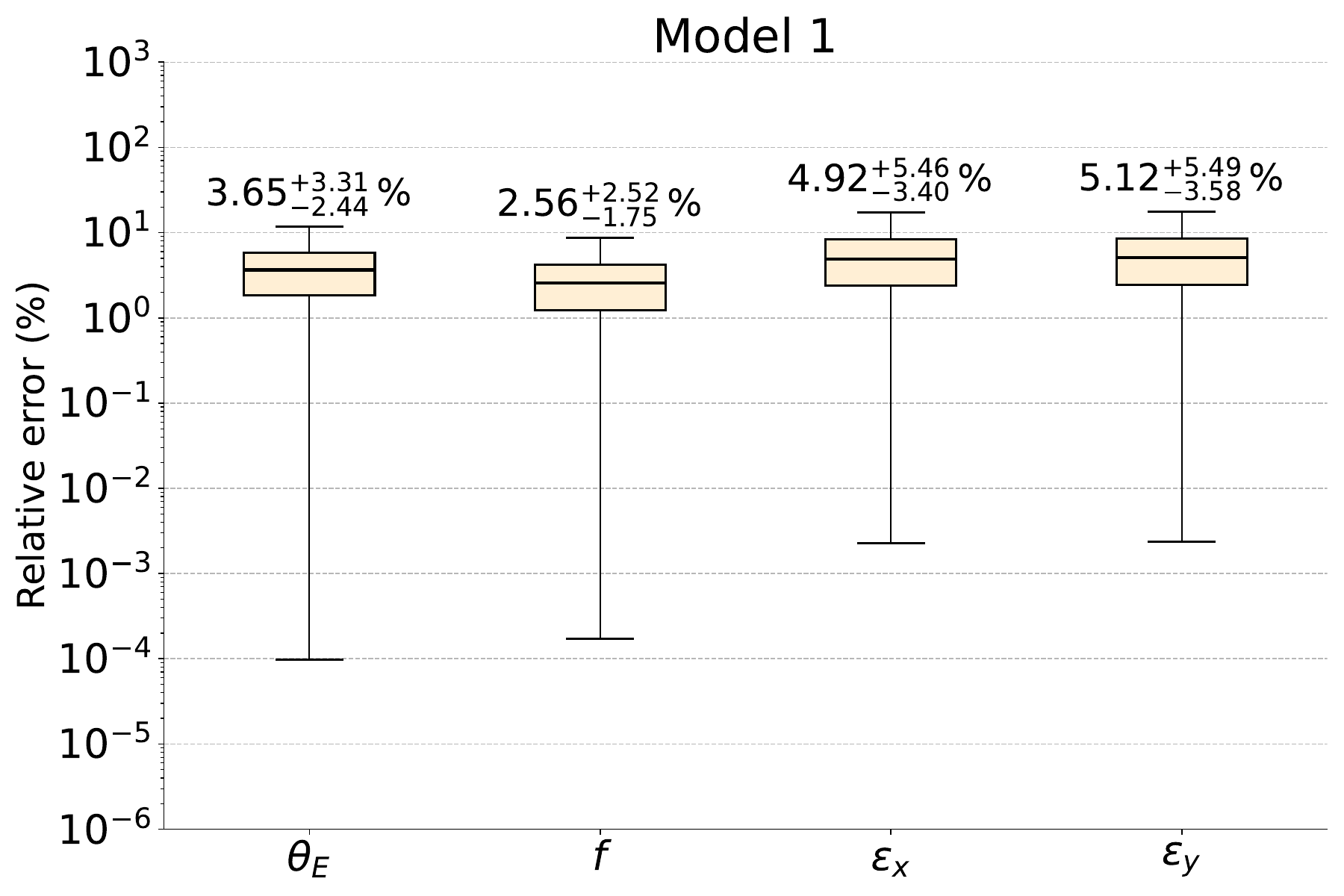}
\includegraphics[width=7cm]{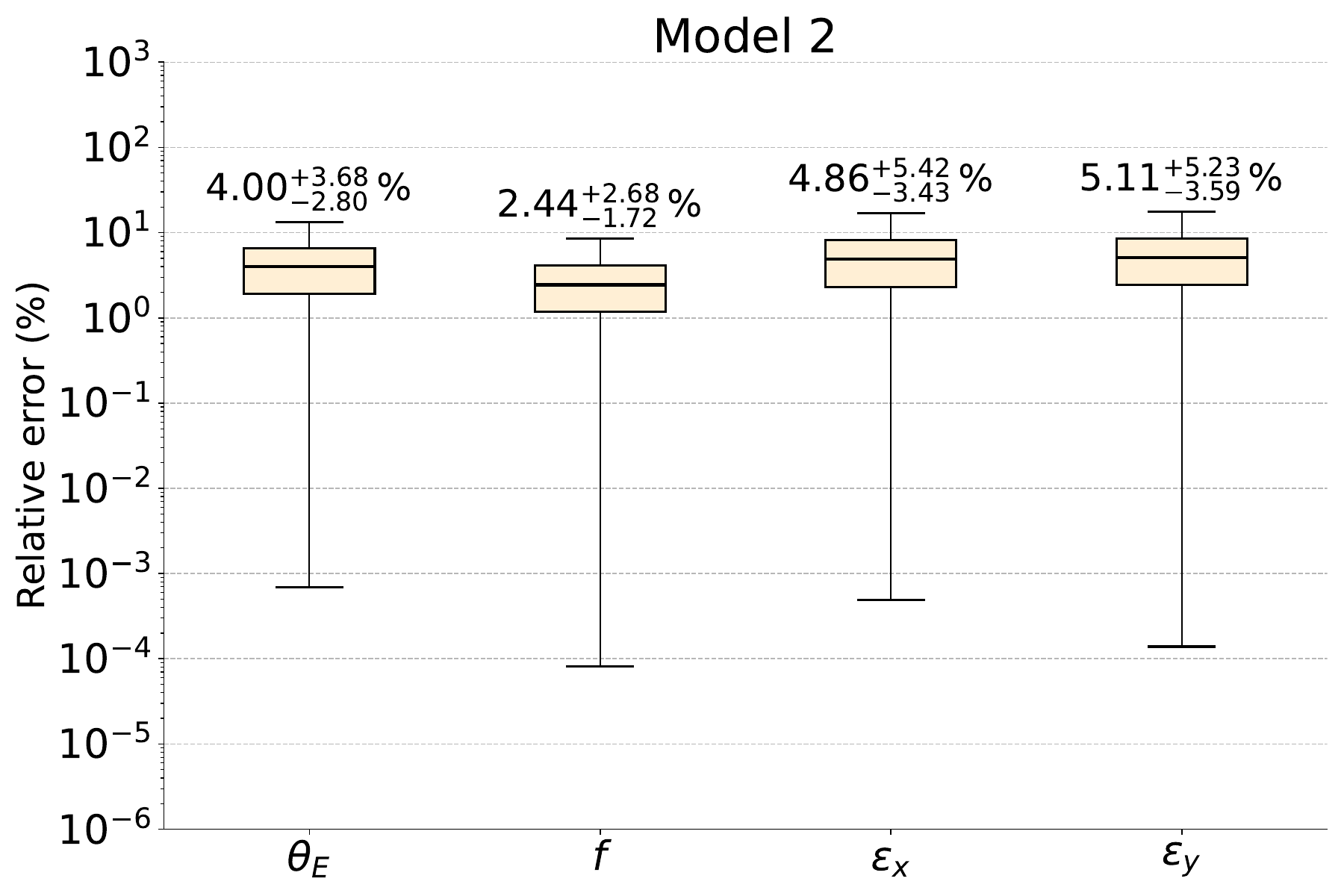}\\
\includegraphics[width=7cm]{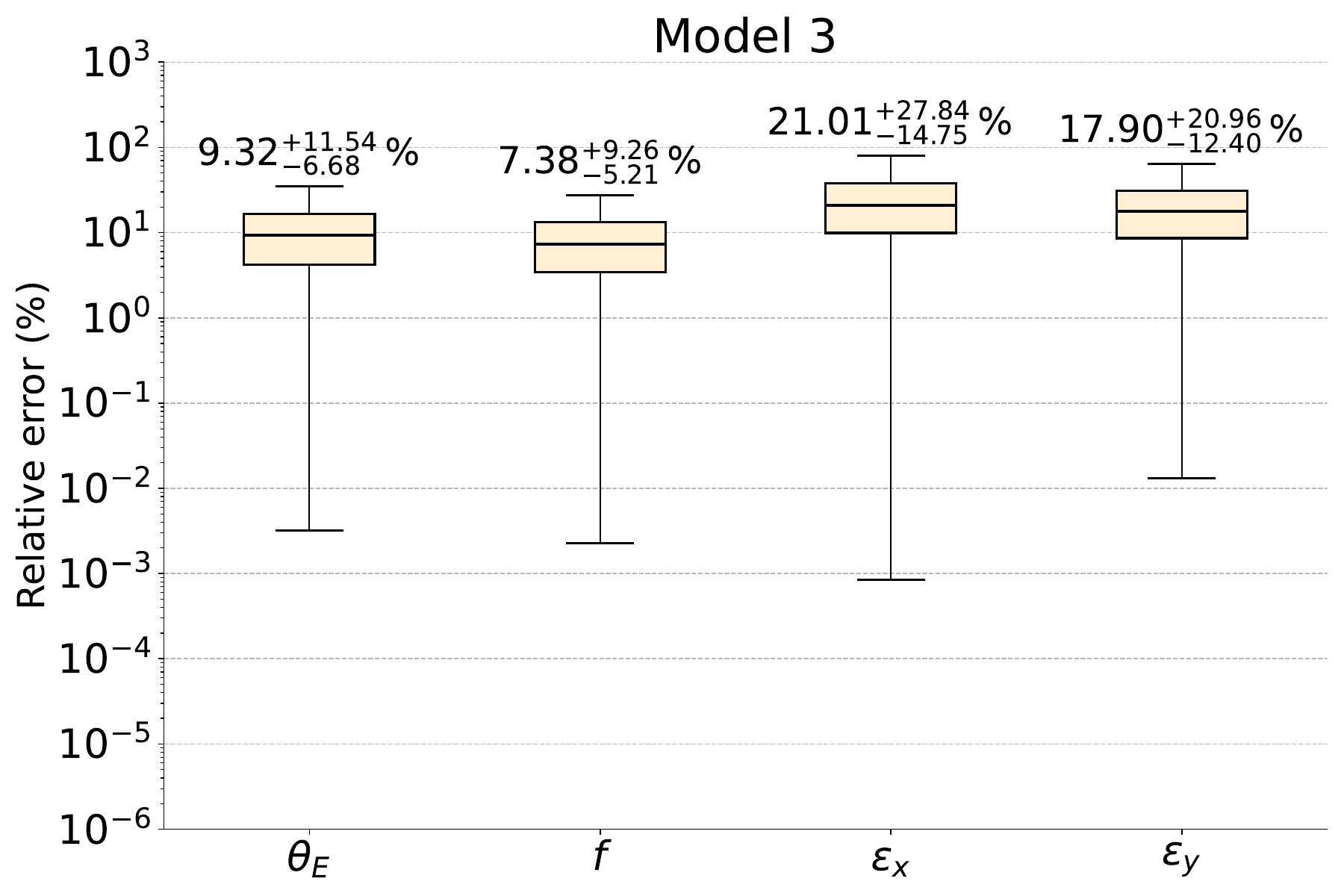}
\caption{Relative errors of the SIE parameters for the three CNN models. Median values and their uncertainty at 68\% confidence level are explicit exposed at the top of the boxplots.}
\label{fig:results_cnn}
\end{figure*}

Finally, the relative errors of the SIE parameters for the three CNN architectures are presented in Figure \ref{fig:results_cnn}. The values shown at the top of each boxplot correspond to the median relative error of the associated parameter. The best predictive performance is obtained for the parameter $f$, with median relative errors as low as $2.6\%$ for the CNN models that incorporate dropout layers; this median error increases by approximately a factor of three when dropout is disabled. The poorest performance is observed for the ellipticity parameters, which exhibit median relative errors up to $5.12\%$ for models 1 and 2, and up to $21\%$ for model 3. These median values are also summarized in Table \ref{table:coefr2}.


\begin{table*}
\centering
\caption{Results for $R^2$ and median $\epsilon_r$ of each parameter for each model. The effect of dropout on generalization is evident. Models 1 and 2 show consistent performance in terms of $R^2$, while the absence of dropout in Model 3 leads to overfitting and reduced generalization capability. Each relative error includes its standard deviation according to the 4-fold-cross procedure}.
\label{table:coefr2}
\renewcommand{\arraystretch}{1.8}
\resizebox{\textwidth}{!}{%
\begin{tabular}{c cc|cc|cc|cc|cc}
\hline
\multirow{2}{*}{\bf Model} 
& \multicolumn{2}{c}{\bf Metrics} 
& \multicolumn{2}{c}{$\boldsymbol{\theta_E}$}
& \multicolumn{2}{c}{$\boldsymbol{f}$}
& \multicolumn{2}{c}{$\boldsymbol{\epsilon_x}$} 
& \multicolumn{2}{c}{$\boldsymbol{\epsilon_y}$} \\
\cline{2-11}
& Loss & MAE 
& $R^2$ & $\epsilon_r$ (\%) 
& $R^2$ & $\epsilon_r$ (\%) 
& $R^2$ & $\epsilon_r$ (\%) 
& $R^2$ & $\epsilon_r$ (\%) \\ 
\hline 
1 & $0.008\pm0.001$ & $0.026\pm0.001$
  & $0.96\pm0.01$ & $3.65^{+3.31}_{-2.44}$ 
  & $0.95\pm0.01$ & $2.56^{+2.52}_{-1.75}$ 
  & $0.96\pm0.01$ & $4.92^{+5.46}_{-3.40}$
  & $0.96\pm0.01$ & $5.12^{+5.49}_{-3.58}$ \\ 

2 & $0.006\pm0.001$ & $0.022\pm0.001$
  & $0.96\pm0.01$ & $4.00^{+3.68}_{-2.80}$ 
  & $0.95\pm0.01$ & $2.44^{+2.68}_{-1.72}$ 
  & $0.97\pm0.01$ & $4.86^{+5.42}_{-3.43}$
  & $0.96\pm0.01$ & $5.11^{+5.23}_{-3.59}$ \\ 

3 & $0.092\pm0.016$ & $0.085\pm0.005$
  & $0.91\pm0.05$ & $9.32^{+11.54}_{-6.68}$ 
  & $0.61\pm0.05$ & $7.38^{+9.26}_{-5.21}$
  & $0.56\pm0.60$ & $21.01^{+27.84}_{-14.75}$
  & $0.64\pm0.15$ & $17.90^{+20.96}_{-12.40}$ \\ 
\hline
\end{tabular}%
}
\end{table*}

\section{Discussion and conclusions}
\label{section:conclusions}

The AlexNet-based CNN architecture was implemented to predict the parameters of the SIE lens model from 76,396 simulated galaxy-galaxy lens systems from the CSST sample \cite{CaoCSST:2024}. The study compared three model configurations: Model 1 added dropout rates of 20\% and 30\% for the first and second dense NN layers; Model 2 used dropout rates of 20\% in both layers; and Model 3 turned off dropout layers to assess their impact on generalization and predictive capabilities. Additionally, the reported uncertainties of the results account for a cross-validation procedure implemented using a 4-fold partitioning scheme.

First, although Model 3 did not exhibit overfitting, both both MSE and MAE error metrics converged to local minima, with values of 0.092 and 0.085, respectively. These are considerably higher than the corresponding minimum values obtained for Models 1 and 2, which are approximately 15 times and 4 times lower for MSE and MAE, respectively. This limitation also affected other performance indicators, such as $R^2$ and PSNR, suggesting that models that incorporate dropout produce higher-quality predictions. We consistently observed that error prediction is substantially improved, i.e., significantly reduced in magnitude, as reported in Table \ref{table:coefr2}. Additionally, we note that Model 1 differs from Model 2 by a 50\% increase in the dropout rate in one of the dense layers. We observe a similar behavior during the training phase and a comparable impact on the final results, which supports the robustness of the proposed technique, and achieving $R^2$ values greater than $0.95$ for the predicted parameters, in contrast to $R^2$ values in the range $0.56-0.91$ obtained with Model 3. Regarding image reconstruction performance, Figure \ref{fig:psnrmeans} reports the PSNR distributions, with median values of 36.9 dB, 36.8 dB, and 29.2 dB for models 1, 2, and 3, respectively. When dropout is incorporated into the models, these values place the reconstructed images in the lower range of the high-quality reconstruction regime. The residual dispersion analysis exposed in Figure \ref{fig:scatter} shows that Models 1 and 2 present negligible systematic bias ($\mu \approx -0.02$) and low NMAD values (0.01-0.04), indicating highly certain and minimal dispersion with a remarkable stability across all morphological parameters. In contrast, while Model 3 has a low average bias ($\mu \approx 0.00$), its high residual dispersion (NMAD 0.07-0.10) limits its reliability for individual parameter inference, indicating a structural inability to capture the data variance.

All three models demonstrated proficiency in predicting the Einstein radius $\theta_E$ and the axial ratio $f$, both of which consistently exhibited the lowest levels of relative errors (Figure \ref{fig:results_cnn}). Models 1 and 2 achieved relative errors with an upper limit around $5 - 9 \%$ in the $90\%$ CL for these two parameters. The parameters related to ellipticity ($\epsilon_x$ ,$\epsilon_y$) proved more difficulty to predict, presenting larger and more dispersed relative errors. This difficulty can be attributed to the degeneracy of the lens equation, where different parameter combinations yield similar images, and the potential confusion of ellipticity components or source asymmetries. Despite these challenges, the overall performance was remarkable, with relative errors around $5-12\%$ at the $90\%$ CL for most SIE parameters. Furthermore, the model struggled to correctly learn the high values for $\epsilon_y$, suggesting a scarcity of training samples with high values or the complexity of interpreting the ellipticity.

The implementation of dropout to enhance gravitational lens studies aligns with previous work, such as Perreault Levasseur et al. \cite{perreaultlUPE:2017}, who optimized this hyperparameter to obtain expected coverage probabilities. Consistent with their findings, this work observed that networks with optimized dropout (Model 2) provided accurate predictions, while the absence of dropout (Model 3) resulted in higher and less reliable predictions. Morningstar et al. Similarly,  \cite{morningstarAIOSG:2018} demonstrated that changing dropout rates affects predictions, achieving good calibration with a dropout rate $1\% $ in a recurrent model. More recently, the LEMON project by Gentile et al. \cite{fabrizioLEMON:2023} and the Euclid Collaboration \cite{euclidQDR:2025} used a Bayesian NN with a dropout rate $3\%$ (keep rate $97\%$) to model lenses for HST and Euclid. Their results, achieving $R^2\sim0.94$ and MAE $\sim0.20$, are comparable to the results obtained in this work ($R^2$ up to 0.97). However, we would remark that these works are performed with different samples and include noise The relevance of the results obtained in Model 2 lies in its balance between computational efficiency and analytical accuracy. While previous lens modeling work using CNN has used dense architectures that require significant resources, our use of an optimized version of AlexNet shows that it is possible to achieve a determination coefficient of $R^2 = 0.97$ with a reduced computational load. Since Einstein's radius ($\theta_E$) is the main indicator of the projected mass within the critical radius, reaching an uncertainty of just 9\% (at 90\% CL) for this parameter has a direct impact on the estimation of the mass enclosed within the $\theta_E$. This precision is vital for cosmology studies to restrict dark matter profiles \cite{binneygalactic:1987, KoopmansETLG:2003}. Therefore, the model not only delivers a quick prediction, but ensures that the derived scientific parameters maintain the integrity necessary for large-scale analysis in missions such as CSST and Euclid. 

Finally, this study confirms that CNNs are promising tools for reducing prediction errors in gravitational lensing modeling. Regularization through dropout is crucial to reduce the variability of these errors and prevent overfitting. The predictions of Models 1 and 2 offered high-quality image reconstruction with a minimal MSE around $2.40 \times 10^{-2}$. Furthermore, this method is significantly faster than traditional Markov Chain Monte Carlo (MCMC) techniques, efficiently predicting SIE parameters using a single GPU. This efficiency facilitates future analyzes of large datasets from modern telescopes such as the Chinese Southern Sky Telescope (CSST) and the Euclid Space Telescope \cite{CaoCSST:2024, gongIntroductionChineseSpace:2026a}. Future research should investigate alternative network architectures—such as ResNet \cite{HeResNet:2016}, ConvNeXt, \cite{LiuConvNet:2022} and U-Net \cite{RonnebergerUnet:2015}—or suitably modified variants thereof, and should incorporate more advanced data augmentation strategies. This would enable a more rigorous characterization of distortions induced by convergence and shear components, particularly in scenarios where the presence of background noise and the finite spatial resolution of the detector could produce image degradation that impacts in the model estimation and significantly complicates their analysis.

\begin{acknowledgments}
We thank anonymous referees for thoughtful remarks and suggestions. J. J. A.-F. thanks to Secretaría de Ciencia, Humanidades, Tecnología e Innovación (SECIHTI) for the master's scholarship support. A.H-.A. acknowledges the support from c\'atedra Marcos Moshinsky (MM), Universidad Iberoamericana for support with the SNI grant and the numerical analysis was also carried out by {\it Numerical Integration for Cosmological Theory and Experiments in High-energy Astrophysics} (Nicte Ha) cluster at IBERO University, acquired through c\'atedra MM support. V.M. acknowledges support from ANID FONDECYT Regular grant number 1231418 and Centro de Astrof\'{\i}sica de Valpara\'{\i}so CIDI 21. J.J.A.-F., A.H.-A., and V.M. acknowledge partial support from project ANID Vinculaci\'on Internacional FOVI240098. 
\end{acknowledgments}

\bibliography{librero0}

\end{document}